\documentclass[a4paper,11pt]{article}
\pdfoutput = 1

\newcommand{\rmi}{\mathrm{i}}
\newcommand{\rme}{\mathrm{e}}
\usepackage{jheppub}
\usepackage{amsmath,amssymb,amscd,braket,amsfonts}
\usepackage{color}
\usepackage[T2A,T1]{fontenc}

\usepackage{graphicx}
\usepackage{slashed}
\usepackage{amsthm}
\usepackage{mathtools}
\usepackage{tikz}
\usepackage[export]{adjustbox}
\usetikzlibrary{calc}
\usetikzlibrary{positioning}
\usetikzlibrary{patterns.meta}
\usepackage{subcaption}
\usepackage{comment}
\usetikzlibrary{decorations.markings}
\usetikzlibrary{patterns.meta}
\usetikzlibrary { decorations.pathmorphing, decorations.pathreplacing, decorations.shapes,calc }

\def\w{{\mathfrak w}}
\def\q{{\mathfrak q}}

\DeclareRobustCommand{\dje}{%
  \ifmmode
    \text{{\fontencoding{T2A}\selectfont\cyrdje}}%
  \else
    {\fontencoding{T2A}\selectfont\cyrdje}%
  \fi
}
\newcommand{\hhbar}{\dje}

\title{Shear viscosity from complex light-cone expansion}
 \author[a]{In\^es Aniceto,}
 \author[a,b]{Pavel Kovtun,}
 \author[a]{Alex Ratcliffe,} 
 \author[c,d]{Micha\l\ Spali\'nski} 
 \affiliation[a]{School of Mathematical Sciences and STAG Research Centre, \\ University of Southampton, Highfield, Southampton SO17 1BJ, UK}
\affiliation[b]{Department of Physics \& Astronomy, University of Victoria, \\ PO Box 1700 STN CSC, Victoria,
BC, V8W 2Y2, Canada}
\affiliation[c]{National Centre for Nuclear Research, 02-093 Warsaw, Poland}
\affiliation[d]{Physics Department, University of Białystok,
15-245 Bia\l ystok, Poland}

 \emailAdd{i.aniceto@soton.ac.uk}\emailAdd{p.kovtun@soton.ac.uk}
 \emailAdd{a.ratcliffe@soton.ac.uk}\emailAdd{michal.spalinski@ncbj.gov.pl}

\abstract{
Near thermal equilibrium, the transverse momentum density in a fluid diffuses according to $\omega(k) \sim - i {\cal D} k^2$ at small $k$, where the diffusion coefficient ${\cal D}$ is proportional to the shear viscosity. In field theories with a holographic dual description in terms of classical gravity, the shear-mode frequency $\omega(k)$ approaches the light cone at large imaginary wave vector. We develop a WKB approach to study this limit. 
For black branes in AdS$_{p+2}$, we 
determine the leading terms of the 
asymptotic large-$k$ expansion of the shear-mode frequency $\omega(k)$.
In the case of AdS$_{5}$, we compute a large number of terms of this series and use resummation techniques to analytically continue the result to small $k$. We find strong evidence that $\omega(k)$ is analytic at $k=0$ and vanishes there, which amounts to the existence of hydrodynamics encoded in the large-$k$ physics. Taking 10 terms in the large-$k$ expansion of $\omega(k)$ gives an answer for ${\cal D}$ which is accurate to better than $0.5\%$.
}

\begin{document}
\maketitle

\section{Introduction}
The emergence of long-distance macroscopic phenomena from short-distance microscopic dynamics is one of the foundational questions in physics. 
Examples include the emergence of rich nuclear phenomena from the simplicity of quarks and gluons at short distances \cite{Peskin:1995ev, Greiner-QCD}, or the emergence of rich condensed-matter phenomena from the simplicity of electrons and nuclei at short distances \cite{Chaikin-Lubensky}. The transition from short-distance simplicity to the long-distance complexity is often difficult to describe quantitatively, and one has to resort to intermediate methods such as phenomenological models or effective field theory.

A particularly instructive case concerns the emergence of classical fluid dynamics at long distances from relativistic quantum field theory at short distances. The intermediate step which is usually invoked in this case is  kinetic theory, however it only applies when the quantum field theory in question is weakly interacting~\cite{Jeon:1995zm, Arnold:2002zm}. While transport coefficients such as shear viscosity can be computed~\cite{Jeon:1994if, Arnold:2000dr}, evaluating the actual spectrum of low-energy excitations near thermal equilibrium from kinetic theory becomes quite non-trivial~\cite{Moore:2018mma}.

Here we are interested in the complementary question: the emergence of classical fluid dynamics from field theories in which the intermediate kinetic-theory description is not applicable. We will focus on strongly interacting field theories which admit a dual holographic description in the framework of the gauge-string correspondence~\cite{Aharony:1999ti}. The literature on hydrodynamics in holography is substantial, starting from the early works on shear viscosity \cite{Policastro:2001yc} and its subsequent universality~\cite{Kovtun:2004de, Buchel:2004qq}, hydrodynamic correlation functions~\cite{Policastro:2002se} and hydrodynamic constitutive relations in the derivative expansion~\cite{Baier:2007ix, Bhattacharyya:2007vjd}. Here, our interest is not in ``deriving'' fluid dynamics from holography per se, but rather in connecting long-distance hydrodynamic physics to short-distance non-hydrodynamic physics. 

In holography, singularities of finite-temperature response functions of the energy-momentum tensor in $(p{+}1)$-dimensional quantum field theory are given by the quasinormal frequencies $\omega(k)$ of the dual asymptotically anti de~Sitter (AdS) black brane in $p{+}2$ dimensions~\cite{Kovtun:2005ev}. Hydrodynamic excitations (shear and sound waves) manifest as special types of ``gapless'' quasinormal modes (QNM) of black branes, whose frequency $\omega$ and the wave vector $k$ satisfy $\omega(k\to0) = 0$. The infinitely many ``gapped'' quasinormal modes with $\omega(k\to0)\neq 0$ describe non-hydrodynamic excitations in field theory. While the gapless small-$\omega$, small-$k$ excitations are hydrodynamic, the large-$k$ excitations are manifestly not, reflecting the short-distance physics of the underlying conformal field theory. At large real $k$, the gapped modes and the sound modes behave as $\omega \sim \pm k + O\left(k^{(1-p)/(3+p)}\right)$~\cite{Festuccia:2008zx, Fuini:2016qsc}. Hydrodynamic shear modes do not approach the light cone at large real~$k$. As we will see in this paper, they do, however, approach the light cone at large imaginary~$k$. This allows one to use the WKB method developed recently in~\cite{Aniceto:2026zmc} to study the large-$k$ asymptotic expansion of the shear-mode frequency $\omega(k)$ at large imaginary $k$.

In related holographic developments, it was realised that hydrodynamic and non-hydrodynamic modes, taken together, form different branches of a multi-valued complex function, opening the possibility to connect small-$k$ modes to large-$k$ modes~\cite{Withers:2018srf, Grozdanov:2019kge, Grozdanov:2019uhi, Grozdanov:2022npo}. Asymptotic large-$k$ expansions of quasinormal frequencies at real~$k$ allow one to probe $k\sim O(\pi T)$ with excellent precision~\cite{Aniceto:2026zmc}. The information contained in quasinormal modes is related to the information contained in the operator product expansion of the dual conformal field theory~\cite{Arnaudo:2026axe, Barrat:2026jfg}, and one may naturally wonder about the relation of hydrodynamics to the short-distance data of the CFT. Thermal stress-tensor correlation functions in holographic CFTs have also been studied using the light-cone OPE, where one finds universal scaling functions in momentum space~\cite{Karlsson:2022osn, Esper:2023jeq}. Here we will study near light-cone singularities of shear-channel correlation functions, at large imaginary momentum.

Our main goal in this paper is to connect the small-$k$ hydrodynamic regime of $\omega(k)$ to the large-$k$ light-cone regime of the same $\omega(k)$. At small $k$, the shear mode has the dispersion relation $\omega(k) = -i {\cal D} k^2 + \dots$, where ${\cal D}$ is the diffusion coefficient of transverse momentum density, ${\cal D} = \eta/(\epsilon+p)$, where $\eta$ is the shear viscosity, $\epsilon$ is the equilibrium energy density, and $p$ is the equilibrium pressure, see e.g.~\cite{Kovtun:2012rj}. In thermal equilibrium with vanishing chemical potentials, $\epsilon + p = Ts$, where $T$ is temperature and $s$ is the density of entropy, hence $\omega = -i \eta/(Ts) k^2 + \dots$. In holography, shear viscosity takes the universal value $\eta/s = 1/4\pi$ in theories with Einstein gravity duals~\cite{Kovtun:2004de} (we use the units in which $\hbar=c=1$). Hence, in any $p\geq2$ we have%
\footnote{
Shear modes do not exist in $p=1$ because there's literally no space for them in 1+1 dimensions. 
}
\begin{align}
\label{eq:shearw-hydro}
  AdS_{p+2}:\ \ \ \ i \omega \sim \frac{1}{4\pi T} k^2  \,,\ \ \ \ k\to0.
\end{align}
Later in the paper, we will present our results in terms of dimensionless quantities
\begin{equation}
\label{eq:wq-def}
    \w \equiv \frac{\omega}{2 \pi T}, \quad \q \equiv  \frac{k}{2\pi T} \,,
\end{equation}
defined so for all $p$.
By rotation invariance, the quasinormal frequencies depend on $\q^2$ rather than just $\q$. Our interest in this paper thus lies in the connection between the hydrodynamic regime and the complex light-cone regime for the shear mode:
\begin{alignat}{2}
\label{eq:wq-1}
   & \text{Hydrodynamics:}\quad & &  \w \sim -i \q^2/2, \ {\rm as}\ \q^2 \to0\,,\\[5pt]
\label{eq:wq-2}
   & \text{Complex light cone:}\quad & &  \w\sim i |\q^2|^{1/2},\ {\rm as}\ \q^2 \to -\infty \,.
\end{alignat}
We will connect the two by determining the asymptotic expansion of the large-$\q$ corrections to \eqref{eq:wq-2}, and then matching the answer to the hydrodynamic expansion at small non-zero~$\q$.

Our presentation is organised as follows. In Section~\ref{sec:shear-overview}, we review the shear mode as a quasinormal mode of the AdS-Schwarzschild black brane in $p{+}2$ dimensions. In Section~\ref{sec:shreq}, we rewrite the master equation for shear-channel perturbations in normal form, which will facilitate the large-$\q$ analysis. We start Section \ref{sec:lc-and-wkb} with an argument from basic quantum mechanics to show that the shear mode approaches the light cone at large imaginary~$\q$. Using this, we review the main ingredients of the WKB method, with particular emphasis on the uniform WKB method for approaching turning points in Section \ref{sec:rev-uniform-WKB}. We put these methods to work in Section~\ref{sec:lc-1}, where we determine the leading correction to the imaginary light-cone scaling for different values of $p$. %
In Section \ref{sec:all_orders}  we turn to the main outcome of this work: evidence of the emergence of hydrodynamics and a computation of the shear viscosity based on the short-distance limit. Focusing on the physically most interesting  case of $p=3$, we use the  uniform WKB approach to develop a prescription for computing  the asymptotic series describing the shear mode frequency to any  order in the large imaginary $\q$ expansion.
In Section \ref{sec:viscosity} we use this expansion together with an analysis based on resurgence and Borel summation to analytically continue the result from large $\q$ all the way to the origin. We find strong evidence that the shear mode frequency is analytic at $\q=0$, and by computing leading terms in the series around that point we calculate the value of the shear viscosity, finding excellent agreement with the known value. 
We discuss our results in Section~\ref{sec:Discussion}.

\section{The shear mode in \texorpdfstring{AdS$_{p+2}$}{AdS p+2}}
\label{sec:shear-overview}

In this section we review the origins of the shear mode as a quasinormal mode of the AdS-Schwarzschild black brane in p+2 dimensions, and set up the main boundary value problem that we will be analysing.

\subsection{Quasinormal modes of the black brane}

We start by reviewing perturbations of $AdS_{p+2}$ black branes. The Einstein equations are $G_{MN} + \Lambda g_{MN} = 0$, where $G$ is the Einstein tensor, $g$ is the metric, $\Lambda = -p(p{+}1)/(2\ell^2)$ is the negative cosmological constant, $\ell$ is AdS radius, and capital indices run from $0$ to $p{+}1$. The black brane metric which solves Einstein's equations can be written in static coordinates as
\begin{align}
\label{eq:bb0}
  ds^2 = \frac{r^2}{\ell^2} \left( - f(r) dt^2 + \sum_{i=1}^p dx_i^2 \right) + \frac{\ell^2}{r^2 f(r)} dr^2 \,,
\end{align}
where $f(r) = 1 - (r_0/r)^{p+1}$. The horizon at $r=r_0$ determines the Hawking temperature $T = r_0 (p{+}1)/(4\pi \ell^2)$. The asymptotic infinity $r\to\infty$ corresponds to the boundary of AdS. In order to study perturbations of the black brane \eqref{eq:bb0}, we write $g_{MN} = \bar g_{MN} + h_{MN}$ where $\bar g_{MN}$ is the background \eqref{eq:bb0}, and linearise Einstein's equations in $h$. By translation and rotation invariance, the perturbations $h_{MN}$ can be taken to be proportional to $\exp(-i\omega t + i k x_p)$, leaving one with a set of coupled ordinary differential equations for $h_{MN}(r)$. These equations separate into three decoupled sets or ``channels'', according to the symmetries of the perturbation under $SO(p{-}1)$ rotations among $x_1, \dots, x_{p-1}$~\cite{Kovtun:2005ev}. The three channels are the ``scalar channel'' (helicity-2), the ``shear channel'' (helicity-1), and the ``sound channel'' (helicity-0).
The shear mode is one of the quasinormal modes in the shear channel; it is the only mode in the shear channel with the property $\omega(k\to0)=0$. 

\begin{figure}
\centering
\includegraphics[width=0.6\textwidth]{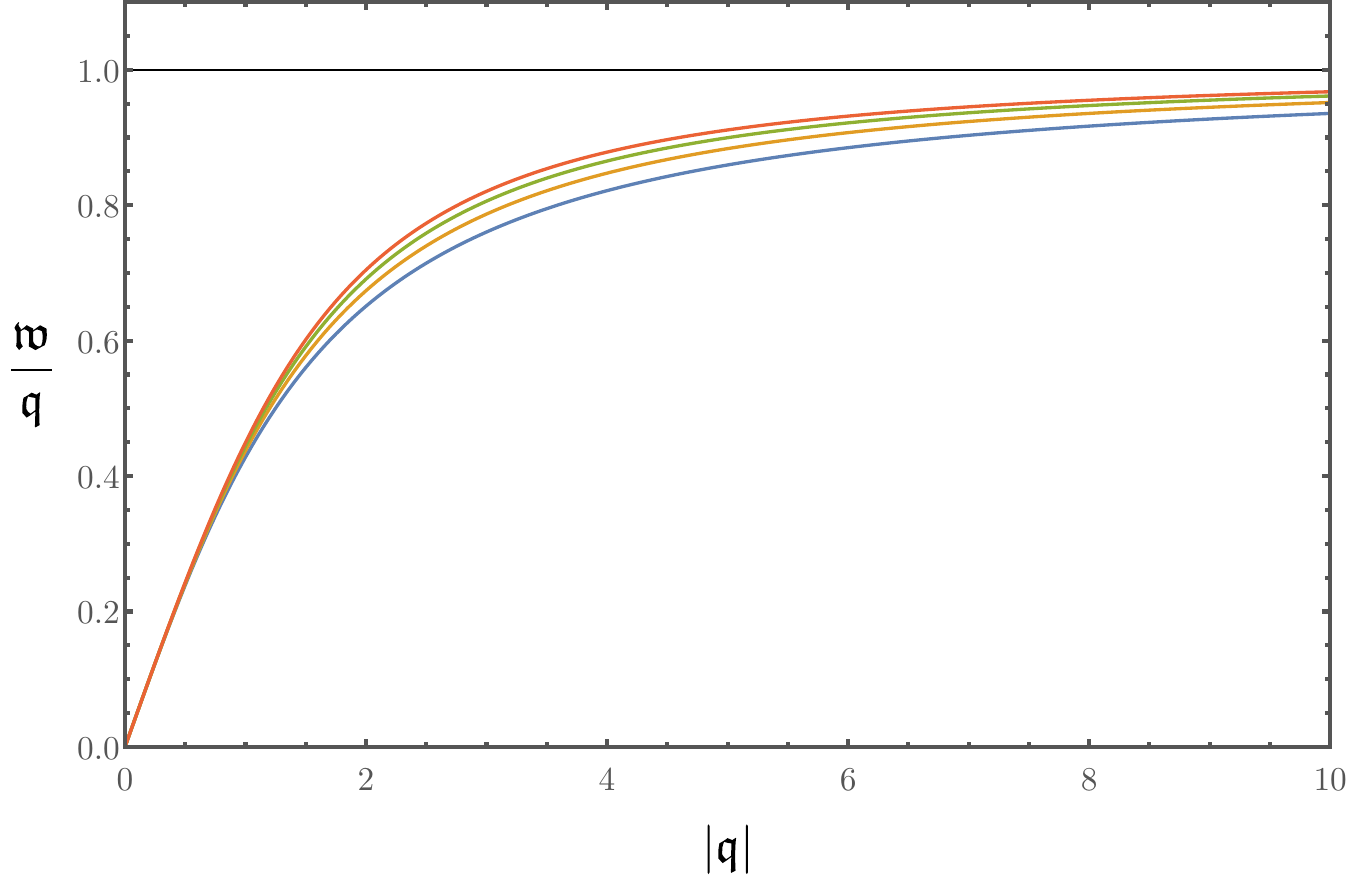}
\caption{
\label{fig:shear-mode-imq}
Shear mode frequency for AdS$_{p+2}$ black branes~\eqref{eq:bb0}, at positive imaginary~$\q$. As $\q^2$ decreases from zero, the shear mode travels up along the imaginary $\w$-axis, and approaches the complex light cone at large imaginary $\q$, i.e.\ $\w\sim\q$, with both $\w$ and $\q$ purely imaginary. The curves correspond to (from bottom to top) $p=2,3,4,5$. The linear behaviour at small $\q$ is a consequence of hydrodynamics. All curves have a slope of $1/2$ at small $\q$, in agreement with eq.~\eqref{eq:wq-1}.
}
\end{figure}

We choose a gauge $h_{rM} = 0$. The shear-channel perturbations are $h_{t x_a}$ and $h_{x_p x_a}$, where $a$ labels one of the coordinates $x_1, \dots, x_{p-1}$. In other words, there are $p{-}1$ identical decoupled shear channels, which in field theory correspond to $p{-}1$ momentum density fluctuations transverse to the chosen spatial wave vector. Following~\cite{Kovtun:2005ev}, we define 
\begin{align}
  Z(r) = \frac{\ell^2}{r^2} \left( k h_{t x_1} + \omega h_{x_{p}x_1} \right) \,.
\end{align}
While the equations for $h_{t x_1}(r)$ and $h_{x_{p}x_1}(r)$ are coupled, the combination $Z(r)$ obeys a single second-order differential equation. 
Introducing the dimensionless $u = r_0/r$ which ranges from 0 (AdS boundary) to 1 (black brane horizon), the ``master equation'' obeyed by shear-channel perturbations is%
\footnote{
The equations for the master variables of~\cite{Kovtun:2005ev} for $p\neq3$ were given in~\cite{Morgan:2009pn}. Note that the coordinate $u = r_0/r$ we use here differs from the coordinate $u$ used in~\cite{Kovtun:2005ev}, and the normalization of $\w$ and $\q$ we use here differs from the normalization used in~\cite{Morgan:2009pn}. 
}
\begin{align}
\label{eq:shear-eqn-d}
  Z''(u) + \left( \frac{\w^2 f'}{f(\w^2 - \q^2 f)}  -\frac{p}{u} \right)  Z'(u) + \frac{(p+1)^2}{4}\frac{\w^2 - \q^2 f}{f^2} Z(u) = 0\,,
\end{align}
where $f(u) = 1 - u^{p+1}$. 
Both $u=0$ and $u=1$ are regular singular points of the equation~\eqref{eq:shear-eqn-d}. 
The indicial exponents near $u=1$ are $\pm i\w/2$, and we choose $- i\w/2$ as the solution corresponding to the ingoing wave at the horizon.
The indicial exponents near $u=0$ are $0$ and $p{+}1$, therefore near the boundary one has
\begin{align}
\label{eq:Z1-bndry}
  Z(u) = {\cal A} (1 + \dots) + {\cal B} u^{p+1} + \dots \,,
\end{align}
where ${\cal A}$ and ${\cal B}$ are the coefficients of the two independent Frobenius-series solutions of eq.~\eqref{eq:shear-eqn-d}. For $Z(u)$ which satisfies the ingoing condition at $u=1$, the near-boundary solution takes the form \eqref{eq:Z1-bndry}, where ${\cal A}$ and ${\cal B}$ depend on $\w$ and $\q$. In fact, as the differential equation depends on $\q^2$, and the boundary conditions are $\q$-independent, the near-boundary coefficients are functions of $\w$ and $\q^2$, reflecting rotation invariance in the boundary field theory. The quasinormal frequencies are defined by the condition
\begin{align}
\label{eq:A=0}
  {\cal A}(\w, \q^2) = 0 \,.
\end{align} 
The solutions to eq.~\eqref{eq:A=0} are the singularities of the retarded two-point function of the helicity-1 components of the energy-momentum tensor in the dual field theory in thermal equilibrium~\cite{Kovtun:2005ev}. One of the solutions of eq.~\eqref{eq:A=0} is the shear mode~\eqref{eq:wq-1}, whose physical meaning is the diffusion of transverse momentum density in the dual field theory. The next small-$\q$ correction to $\w(\q)$ can be obtained too~\cite{Baier:2007ix, Kapusta:2008ng},
\begin{align}
  \w = -\frac{i}{2} \q^2 - \frac{i}{8} \q^4 \, (p+1) \!\int_0^1 \frac{u^p (1-u^2)}{1-u^{p+1}} du  + O(\q^6) \,.
\end{align}
The integral can be expressed in terms of the digamma function. Physically, the coefficient of the $O(\q^4)$ term is related to higher-order hydrodynamic transport coefficients in the dual field theory. Higher-order terms in the small-$\q$ expansion for $p=3$ can be found in~\cite{ Grozdanov:2019uhi, Arnaudo:2026axr}. As $\q^2$ decreases from zero, the shear-mode frequency travels up the imaginary axis, and eventually enters the light-cone regime~\eqref{eq:wq-2}. Figure~\ref{fig:shear-mode-imq} shows numerical results for the shear-mode $\w(\q)$ at $\q^2<0$, illustrating the asymptotic behaviour $\w \sim \pm\q$, as $\q\to\pm i \infty$.

\subsection{The master equation in normal form}
\label{sec:shreq}

Equation~\eqref{eq:shear-eqn-d} has a transparent physical interpretation, allowing one to access both the quasinormal modes and the actual response functions of the dual field theory~\cite{Kovtun:2005ev}. However, from a technical standpoint, the presence of $(\w^2 - \q^2 f)$ in the denominator produces $\w$- and $\q$-dependent singular points in the differential equation, which may be inconvenient for analytical calculations of the quasinormal spectrum. One can perform a variable change in eq.~\eqref{eq:shear-eqn-d} which gets rid of the pesky singular points, leaving one with the differential equation whose coefficients do not contain $\w$ and $\q$ in the denominator. Indeed, suppose we have an equation of the form
\begin{align}
  Z''(u) + \left( Q - \frac{R'}{R} \right) Z'(u) + R S Z(u) = 0 \,.
\end{align}
Variable change $Z'(u) = R(u) X(u)$ gives rise to
\begin{align}
  X''(u) + \left( Q - \frac{S'}{S} \right) X'(u) + \left( RS + Q' - Q \frac{S'}{S} \right) X(u) = 0\,,
\end{align}
swapping singular points due to $R(u)$ in the equation for $Z$ for singular points due to $S(u)$ in the equation for $X$. Applying this to eq.~\eqref{eq:shear-eqn-d} with $R(u) = \w^2 - \q^2 f(u)$, one finds
\begin{align} 
\label{eq:psi2eqn}
  X''(u) + \left( -\frac{p}{u} + \frac{3f'}{f}\right) X'(u) + \left( -\frac{p^2}{u^2} + (p+1)^2\frac{\w^2 - \q^2 f}{4f^2} + \frac{p(p{+}1)}{u^2 f} + \frac{f'^{\,2}}{f^2}\right) X(u) = 0\,.
\end{align}
Of course, there are many ways to write down equations which would not have the frequency in the denominator. Taking $X(u)= \Phi(u) u^{p/2}/f(u)$ in eq.~\eqref{eq:psi2eqn}, one finds
\begin{align}
  \label{eq:Phieqn}
  \Phi''(u) + \frac{f'}{f}\Phi'(u) + \left( (p+1)^2\frac{\w^2 - \q^2 f}{4f^2} - \frac{p f'}{2u f} - \frac{p(p{-}2)}{4u^2} \right) \Phi(u) = 0\,.
\end{align}
This equation was written down in ref.~\cite{Morgan:2009pn} (there is an overall minus sign there in the definition of $\Phi$ compared to ours), following the analysis of gravitational variables in~\cite{Kodama:2003jz}. Similar to $Z(u)$, the near-horizon exponents for $\Phi(u)$ are $\pm i\w/2$. The near-boundary exponents for $\Phi$ are $1-p/2$ and $p/2$, hence the small-$u$ expansion is
\begin{align}
\label{eq:Phi-bndry}
  \Phi(u) = {\cal A}_{\Phi} \, u^{1-\frac{p}{2}} (1 + \dots) + {\cal B}_{\Phi} u^{\frac{p}{2} } + \dots \,, 
\end{align}
where ${\cal A}_\Phi$ and ${\cal B}_\Phi$ are the coefficients of the two independent Frobenius-series solutions of eq.~\eqref{eq:Phieqn}. 
As was pointed out in ref.~\cite{Morgan:2009pn}, the coefficients ${\cal A}$ and ${\cal A}_\Phi$ in eqs.~\eqref{eq:Z1-bndry} and \eqref{eq:Phi-bndry} are related by ${\cal A}_\Phi = {\cal A}/(p{-}1)$, hence the condition ${\cal A} = 0$ is equivalent to the condition ${\cal A}_\Phi = 0$. Consequently, the correct quasinormal spectrum \eqref{eq:A=0} can be obtained by solving eq.~\eqref{eq:Phieqn} with the ingoing condition at the horizon $u=1$, and the Dirichlet condition at the boundary $u=0$. 
The advantage of using eq.~\eqref{eq:Phieqn} is that the factor $\w^2$ appears linearly in the equation, which makes it a convenient starting point for the WKB analysis.

Eliminating the first derivative from eq.~\eqref{eq:Phieqn} through $\phi(u) = f(u)^{1/2} \Phi(u)$, the equation can be written in a normal form,
\begin{align}\label{eq:Srchodinger-eq-u}
  \phi''(u) - V(u) \phi(u) = 0\,,\
\end{align}
where the potential is 
\begin{align}
  V(u) = -\frac{(p+1)^2\w^2}{4f^2} +\frac{(p+1)^2\q^2}{4f} - \frac{(p{+}1)^2 u^{2p}}{4f^2} - \frac{p(p{+}1) u^{p-1}}{f} + \frac{p(p{-}2)}{4u^2}  \,,
\end{align} 
This is the form of the equation governing the shear channel perturbations that will be the basis of the WKB analysis presented in the following. 
The boundary conditions that quantise the QNM frequencies are
\begin{equation}\label{eq:bc_u}
    \phi(u) \sim \begin{cases}
        u^{p/2}, & \quad u \to 0,\\
        (1-u)^{1/2 -\rmi \w /2}, &\quad u\to 1.
    \end{cases}
\end{equation}
When $\w$ is positive imaginary, these become Dirichlet boundary conditions.


\section{Light-cone scaling and WKB}
\label{sec:lc-and-wkb}

We now introduce the asymptotic limit which is the focus of this paper: a complex light-cone scaling. We then transform the QNM equation into a form amenable to further treatment. To effectively solve this equation we will use the WKB method, which we review later in this section. The methods hereby presented will then be used in the next sections to determine the sub-leading corrections to the complex light-cone scaling for general dimension $p\ge2$ in Sec.~\ref{sec:lc-1}, and then the large $\q$ expansion of the shear mode frequency for $p=3$ in Sec.~\ref{sec:all_orders}.

\subsection{Light-cone scaling}
\label{sec:lc-scaling}

We are interested in QNM where $\w$ is positive imaginary and $\q$ is imaginary. The light-cone scaling of the corresponding QNM can be seen from quantum-mechanical arguments, analogous to what was done recently in \cite{Buchel:2026rep, Buchel:2026ajv} in settings with higher-derivative gravity. Starting with eq.~\eqref{eq:Phieqn}, we introduce a new coordinate $y$, such that $dy/du = 1/f(u)$, and choose $y(u{=}0)=0$. The coordinate $y$ is a monotonic function of $u$ ranging from $0$ to $\infty$, with
\begin{align}
    y = u + \frac{u^{p+2}}{p+2} + O(u^{2p+3}) \,.
\end{align}
Next, we define $\hhbar^2 \equiv -1/\q^2 > 0$ and $E\equiv -\frac{(p+1)^2}{4}\frac{\w^2}{\q^2} < 0$, so that eq.~\eqref{eq:Phieqn} takes the form of Schr\"odinger's equation:
\begin{align}
\label{eq:Schreq-Phi}
  -\hhbar^2\frac{d^2 {\Phi}}{dy^2} + U(y)\, {\Phi}(y) = {E}\,{\Phi}(y) \,,
\end{align}
where the potential is $U(y) = U^{(0)}(y) + \hhbar^2 U^{(2)}(y)$,
\begin{align}
    U^{(0)}(y) = -\frac{(p+1)^2}{4} f(u(y)) \,,\ \ \ \ 
    U^{(2)}(y) = f(u(y))\left[\frac{p(p-2)}{4u(y)^2} - \frac{3p^2}{4}\,u(y)^{p-1}\right]\,.
\end{align}
With $\w$ positive imaginary, the boundary conditions obeyed by $\Phi$ are still Dirichlet boundary conditions at $u=0$ and $u=1$, and the QNM problem becomes the standard quantum-mechanical problem of bound states in an attractive potential. In the classical limit $\hhbar\to0$, the energy of each fixed bound state will approach the minimum value of the classical potential. With $f(u)=1-u^{p+1}$, the minimum of $U^{(0)}$ is at $u=0$, which corresponds to
\begin{equation}
    \w^2=\q^2\,.
\end{equation}
This motivates setting 
\begin{equation}
\label{eq:lcscaling}
    \w= \q \,s(\q)\,,
\end{equation}
with $s \sim \pm 1$ as $|\q|\to\infty$; from now on we consider the ``$+$'' sign without loss of generality. 

\begin{figure}
\centering
\includegraphics[width=0.6\textwidth]{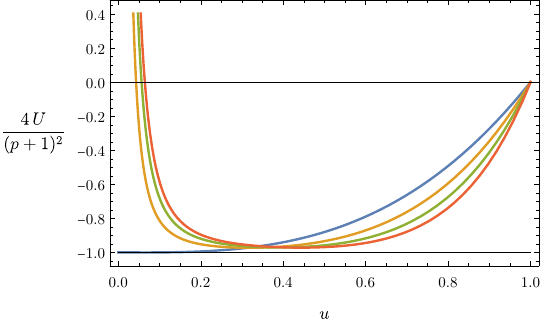}
\caption{
\label{fig:potentials}
The potential $U(y)$ in eq.~\eqref{eq:Schreq-Phi}, plotted in the coordinate $u$, and rescaled by a factor of $4/(p{+}1)^2$, for $\hhbar=1/10$. The curves correspond to (from bottom to top at small~$u$) $p=2,3,4,5$. As $\hhbar\to0$, all rescaled potentials approach $-1+u^{p+1}$, and the bound state wave functions are localized near $u=0$. Thin horizontal lines mark $0$ and $-1$.
}
\end{figure}

The potential $U(u)$ is shown in Fig.~\ref{fig:potentials}.
At small $\hhbar$, the wave functions of bound states are localized near $y=0$. The potential at small $y$ is 
\begin{align}
\label{eq:U-small-y}
    U(y) = -\frac{(p{+}1)^2}{4} + \frac{(p{+}1)^2}{4} y^{p+1} + \hhbar^2 \frac{p(p{-}2)}{4y^2} + \dots \,.
\end{align}
Move the first term to the right-hand side of eq.~\eqref{eq:Schreq-Phi}, and consider a bound-state wave function of width $l$ localized near $y=0$. The kinetic energy scales as $\hhbar^2/l^2$, and the potential energy scales as $l^{p+1}$; in order for the two contributions to balance each other, $l$ has to scale as $\hhbar^{2/(p+3)}$. This scaling makes the second and the third terms in the small-$y$ expansion in \eqref{eq:U-small-y} of the same order (when $p=2$, the second term balances the kinetic energy), with the dropped terms subleading at small~$y$. As a result, $E = -(p{+}1)^2/4 + O(\hhbar^{2(p+1)/(p+3)})$, or
\begin{align}
\label{eq:wq-lc-1}
   \frac{\w^2}{\q^2} = 1 + O\!\left( \q^{-\frac{2(p+1)}{(p+3)}} \right)\,.
\end{align}
For imaginary $\q$, the correction in eq.~\eqref{eq:wq-lc-1} is negative. This is immediate for $p\geq3$ because finite-$\hhbar$ contributions at small $\hhbar$ shift the potential upwards from $-(p{+}1)^2/4$, as can be seen in Fig.~\ref{fig:potentials}. Determining the sign of the correction requires more care at $p=2$ because in that case $U(y)$ develops a shallow minimum below $-(p{+}1)^2/4$ at small~$\hhbar$. Nevertheless, the sign of the correction in eq.~\eqref{eq:wq-lc-1} for $p=2$ is still negative because the zero-point energy overcomes the shallow dip in $U(y)$.%
\footnote{
   At $p=2$, the minimum of $U$ is located at $u_{\rm min} = \frac23 \hhbar + O(\hhbar^4)$, which gives $U_{\rm min} = -\frac94 - \frac43 \hhbar^3 + O(\hhbar^6)$. The shallow depth makes the dip unnoticeable in Fig.~\ref{fig:potentials}. The small-$y$ potential is $U(y) = -\frac94 + \frac94 y^3 - 3\hhbar^2 y + \dots$. For a wave function of width $l = O(\hhbar^{2/5})$, the contributions from the kinetic energy and the $O(y^3)$ terms in the potential which scale as $O(\hhbar^{6/5})$ overcome the $O(\hhbar^{12/5})$ contribution from the $O(\hhbar^2 y)$ term which is responsible for the shallow dip. 
} 
In fact, the sign  of the correction in eq.~\eqref{eq:wq-lc-1} is demanded by causality of the boundary theory, which requires analyticity of the retarded function for ${\rm Im}(\w) > |{\rm Im}(\q)|$. As QNMs are poles of the retarded function, this translates to the requirement that the QNMs must obey ${\rm Im}(\w) \leq |{\rm Im}(\q)|$~\cite{Heller:2022ejw}. Correspondingly, the function $s(\q)$ in eq.~\eqref{eq:lcscaling} must be such that $|s(\q)|\leq 1$. 

The coefficient of the correction in eq.~\eqref{eq:wq-lc-1} can be calculated as follows. 
Motivated by the scaling $l\sim \hhbar^{2/(p+3)}$, introduce a rescaled coordinate $\tau$ by $y=\hhbar^{\frac{2}{p+3}} \,\tau$, which allows to zoom in on the small-$y$ ``boundary-layer'' region where the bound states are localized, and the associated small parameter $\varepsilon = \hhbar^{2(p+1)/(p+3)}$. Writing the energy as $E = -\frac{(p+1)^2}{4} + \varepsilon E^{(1)}$, the Schr\"odinger equation with the leading-order potential \eqref{eq:U-small-y} becomes
\begin{align}
\label{eq:Schreq-BL}
  -\frac{d^2 \Phi}{d\tau^2} + \left( \frac{p(p{-}2)}{4\tau^2} + \frac{(p{+}1)^2}{4} \tau^{p+1} \right) \Phi = E^{(1)} \Phi \,,
\end{align}
again with Dirichlet boundary conditions at $\tau=0$ and $\tau=\infty$. 
If we write the correction to the QNM frequency as
\begin{align}
\label{eq:wq-lc-2}
   \frac{\w^2}{\q^2} = 1 - k_0  \, |\q|^{-\frac{2(p+1)}{(p+3)}} + \dots\,,
\end{align}
this gives $k_0 = 4E^{(1)}/(p{+}1)^2$.  Note that the potential in eq.~\eqref{eq:Schreq-BL} is such that $E^{(1)}>0$, hence $k_0>0$, as it should be. The ground state energy 
in eq.~\eqref{eq:Schreq-BL} can be determined numerically, and one finds 
\begin{equation}\label{eq:k_0_num}
\begin{aligned}
    & p=2: \ \  k_0 \approx 2.1212\,,\\
    & p=3: \ \  k_0 \approx 2.1407\,, \\
    & p=4: \ \  k_0 \approx 2.1196\,,\\
    & p=5: \ \  k_0 \approx 2.0827\,.
\end{aligned}
\end{equation}
At large imaginary $\q$, the shear mode has the largest value of positive imaginary $\w$, i.e.\ it is the closest mode to the complex light cone $\w\sim\q$. In the quantum-mechanical problem, the shear mode therefore corresponds to the ground state of the problem~\eqref{eq:Schreq-Phi}, which translates to the ground state of the problem~\eqref{eq:Schreq-BL}. Thus, eq.~\eqref{eq:wq-lc-2} with the numerical values of $k_0$ in eq.~\eqref{eq:k_0_num} provides the leading large-$|\q|$ asymptotics for the curves shown in Fig.~\ref{fig:shear-mode-imq}. 

Later in the paper, we will develop a more systematic method for evaluating not only $k_0$, but all further small-$\varepsilon$ corrections to $\w^2/\q^2$, the ``$\dots$'' in eq.~\eqref{eq:wq-lc-2}, which does not rely on self-adjointness of the eigenvalue problem. In the following sections, we will use the normal form \eqref{eq:Srchodinger-eq-u} introduced in Section \ref{sec:shear-overview} instead of eq.~\eqref{eq:Schreq-Phi} analysed here, as the potential takes a simpler analytic form when expressed in terms of $u$.

\subsection{A brief review of the WKB method} 
\label{sec:rev-WKB}

We now turn to the main tools we will need to study the behaviour of the frequency $\w(\q)$ beyond the leading complex light-cone scaling. 
We begin with a brief review of how to construct global solutions (in $u$) to the shear mode equation \eqref{eq:Srchodinger-eq-u} using the exact WKB method \cite{AIHPA_1983__39_3_211_0,delabaere:hal-01886535,Delabaere1997ExactSE}. See \cite{kawai2005algebraic,Iwaki:2014vad} for more detailed reviews.  This method enables one to derive exact quantisation conditions (EQCs) which capture \emph{all} perturbative and non-perturbative contributions to the energy or, in our case, the QNM frequency.

Taking the light cone scaling \eqref{eq:lcscaling}, the QNM equation \eqref{eq:Srchodinger-eq-u} at large $\q$ becomes a singularly perturbed equation with potential
\begin{equation}
    V(u)=\q^2 \, V_0(u)+V_2(u)\,,
\end{equation}
where
\begin{align}
    V_0(u)&= -\frac{(p+1)^2\,s^2}{4f^2} +\frac{(p+1)^2}{4f}\,\\
    V_2(u)&=- \frac{(p{+}1)^2 u^{2p}}{4f^2} - \frac{p(p{+}1) u^{p-1}}{f} + \frac{p(p{-}2)}{4u^2}\,.
\end{align}
To analyse this equation  in the large-$\q$ regime we will use the WKB method.

The WKB method starts with the following ansatz for $\phi(u)$:
\begin{equation}
\label{WKBansatz}
\phi(u)= \exp\left( \int^u S(u') \,du'\right)\, .
\end{equation}
The function $S(u)$ obeys the Riccati equation 
\begin{equation} 
\label{eq:Riccati}
    S'(u) +S^2(u) -\q^2\,V_0(u) -V_2(u) =0\,.
\end{equation}
Assuming that $|\q|$ is large, we expand $S(u)$ perturbatively in $1/\q$ as 
\begin{equation}\label{wkbSexpansion}
S(u) = \sum_{n=-1}^\infty \q^{-n} S_n(u).
\end{equation}
Using the large-$\q$ expansion of $S(u)$ in the Riccati equation \eqref{eq:Riccati}, one finds that the components $S_n(u)$ obey
\begin{equation}
\label{eq:riccati_perturb}
\begin{aligned}
S_{-1}(u)^2 &=V_0(u)\\
S_0(u)&=-\frac{S_{-1}'(u)}{2S_{-1}(u)},\\
S_{1}(u) &= -\frac{1}{2S_{-1}(u)}\left( S_0'(u)+S_{0}(u)^2 -V_{2}(u)\right),\\
S_{j+1}(u) &= -\frac{1}{2S_{-1}(u)} \left( S_j'(u) + \sum_{k=0}^j S_{j-k}(u) S_k(u)\right), \quad j\ge 1.
\end{aligned}
\end{equation}
Solving at leading order, we have
\begin{equation}
    S_{-1}(u) = \pm\sqrt{V_0(u)}
\end{equation}
and we define two formal solutions $S_+(u)$ and $S_-(u)$ to \eqref{eq:riccati_perturb}, according to whether the recursion is solved starting with the plus or minus sign respectively. A useful linear combination  of the two solutions is
\begin{equation}\label{eq:sodd_def}
    S_{\text{odd}}(u) = \frac{1}{2}\left(S_+(u) -S_-(u)\right), \quad  S_{\text{even}}(u) = \frac{1}{2}\left(S_+(u) +S_-(u)\right)\,  ,
\end{equation}
as using the Riccati equation \eqref{eq:Riccati} one can show that $S_{\text{even}}(u) =-\frac{1}{2}\frac{d}{du} \ln S_{\text{odd}}(u)$. This implies that the basis of formal solutions can be expressed solely through $S_\text{odd}$:
\begin{equation} \label{eq:WKB_solutions}
    \phi_\pm(u) = \frac{1}{\sqrt{S_{\text{odd}}(u)}}\exp\left(\pm\int^u S_{\text{odd}}(u')\, du'\right)\,,
\end{equation}
where for the moment, we do not specify the lower limit of the integral. Note that $S_{\text{odd},-1}(u)=\sqrt{V_0(u)}$.

When expanded for large-$\q$, the objects defined by \eqref{eq:WKB_solutions} are typically divergent, formal series in $\q^{-1}$ (after the leading exponential containing $\q\,S_{-1}$ has been isolated). The actual analytic solutions to \eqref{eq:Srchodinger-eq-u} can be reconstructed from the formal series \eqref{eq:WKB_solutions} using the machinery of resummation and resurgence (see \cite{sauzin2014introduction1summabilityresurgence, Aniceto_2019, Dorigoni_2019} for reviews on the subject). However, we choose to work at the level of formal series and acknowledge that a fixed linear combination of formal solutions \eqref{eq:WKB_solutions} will not be sufficient to globally reconstruct a choice of analytic solution to \eqref{eq:Srchodinger-eq-u} for all $u$. 
If we let $F(u)$ denote a choice of analytic solution to \eqref{eq:Srchodinger-eq-u}, then in one region of the complex plane we may have
\begin{equation}
    F(u) \sim_{\q \to \infty} A\phi_+(u) + B\phi_-(u)
\end{equation}
but in another region of the complex-$u$ plane, we may have
\begin{equation}
    F(u) \sim_{\q \to \infty} (A+C)\phi_+(u) + (B+D)\phi_-(u)
\end{equation}
where $A, B, C, D$ are constants and by $\sim_{\q \to \infty}$ we mean that the formal series on the RHS can be used to reconstruct the LHS exactly. The discontinuous jumps in the large-$\q$ series representing the actual analytic function are known as \emph{Stokes jumps}. 

Stokes jumps always occur when one solution is maximally exponentially enhanced with respect to the other. In WKB analysis, this occurs when
\begin{equation} 
\label{eq:SG_condition}
   \text{Im}\, \left(\q \int_{u^*}^u \sqrt{V_0(u')}\,du'\right) = 0\,,
\end{equation}
where the points $u^*$ are the poles of $S_\text{odd,j}$ for all late $j$ known as \emph{turning points}.  From the recurrence relations  \eqref{eq:riccati_perturb}, it is clear that any zeros or simple poles of $V_0(u)$ will be turning points. This equation defines lines in the complex $u$-plane that separate regions where a given linear combination of formal solutions $\phi_\pm(u)$ may reconstruct a chosen analytic solution to \eqref{eq:Srchodinger-eq-u}. These lines in the complex-$u$ plane are called \emph{Stokes lines}. The set of all Stokes lines is called the \emph{Stokes graph}. The relation \eqref{eq:SG_condition} implies that along a Stokes line, the quantity 
\begin{equation} 
\label{eq:SG_condition_real}
   \text{Re}\; \left(\q \int_{u^*}^u \sqrt{V_0(u')}\,du'\right) 
\end{equation}
has a definite sign. Stokes lines for which this quantity is positive/negative are called \emph{positive/negative Stokes lines}. The sign of \eqref{eq:SG_condition_real} determines whether $\phi_+(u)$ or $\phi_-(u)$ is exponentially enhanced relative to the other one.

\begin{figure}[!h]
    \centering
   \begin{tikzpicture}[scale=3]

\draw[] (-0.5, 0.866)--(0, 0);
    \draw[] (-0.5, -0.866)--(0, 0);
       \draw[] (1, 0)--(0, 0);
    \draw[decorate,blue,decoration={zigzag, segment length = 2mm,amplitude=0.5mm}] (-1, 0)--(0, 0);
       \node[] at (-0.4, 0.866) {$-$};
        \node[] at (-0.4,-0.866) {$-$};
         \node[] at (1,0.1) {$+$};
             \node[] at (0.9,0.9) {II};
              \node[] at (0.9,-0.9) {I};
               \node at (0.5,-0.5) {$\phi^{I}_{\pm}$};
                \node at (0.5,+0.5) {$\phi^{II}_{\pm}$};
                 \node (A) at (0.5-0.1,0.5-0.15) {$\times$};
              \node (B) at (0.5-0.1,-0.5-0.15) {$\times$};
  \draw[->, thick, bend left=30] (B) to (A);

       \filldraw[red] (0, 0) circle (1pt) node[black,above right] {$u^*$};
   \end{tikzpicture}
    \caption{Diagram depicting the analytic  continuation and Stokes jump at the turning point $u^*$ with $V_0(u) \propto(u-u^*) + \mathcal{O}\left((u-u^*)^2\right)$. $\phi^I_\pm$/$\phi^{II}_\pm$ is the basis of WKB solutions defined in the regions $I$/$II$. The solid black lines represent Stokes lines and the blue zig-zagged line represents the branch cut. The labels $+/-$ denote the sign of the Stokes lines. We emphasise that not every turning point has an associated branch cut.}
    \label{fig:Stokes_Jump}
\end{figure}
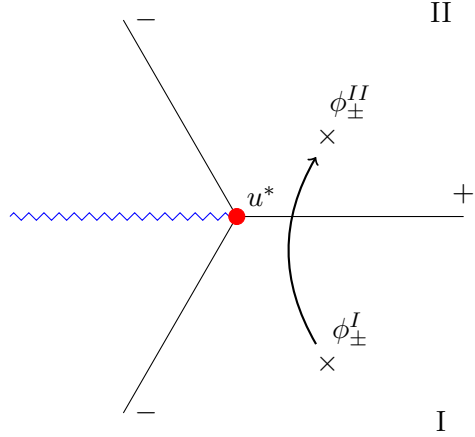

We can now express how the Stokes jumps occur more precisely. Let $I$ and $II$ be two regions on either side of a Stokes line emanating from the turning point $u^*$. Then let $\phi^I_{\pm}(u)$ be the formal large-$\q$ solutions defined in  region $I$, and $\phi^{II}_{\pm}(u)$ be formal large-$\q$ solutions defined in region $II$, see Figure \ref{fig:Stokes_Jump}. In general, the analytic continuation of $\phi_\pm^I(u)$ from the region $I$ to the region $II$ across a Stokes line emanating from $u^*$ in the counter-clockwise direction, can be expressed by the following matrix equations%
 \begin{equation}\label{eq:stokes_jump}
     \begin{cases}
        \left( \begin{matrix}
         \phi^I_+\\
         \phi^I_-
     \end{matrix}\right) = \left(\begin{matrix}
         1& A(\q) \\0&1
     \end{matrix} \right)\left( \begin{matrix}
         \phi^{II}_+\\
         \phi_-^{II}
     \end{matrix}\right), \quad \text{Re}\,\left( \q  \int^{u}_{u^*} \sqrt{V_0(u')} \, du'\right) > 0\,;\\[15pt]
         \left( \begin{matrix}
         \phi^I_+\\
         \phi^I_-
     \end{matrix}\right) = \left(\begin{matrix}
         1& 0 \\B(\q)&1
     \end{matrix} \right)\left( \begin{matrix}
         \phi^{II}_+\\
         \phi^{II}_-
     \end{matrix}\right), \quad \text{Re}\,\left(\q \int^{u}_{u^*} \sqrt{V_0(u')} \, du'\right) < 0 \,,\\
    \end{cases} 
 \end{equation}
for some $u$-independent parameters $A(\q)$ and $B(\q)$ called \emph{Stokes constants}. Generically, these constants can be functions of the large parameter $\q$ and will depend on the choice of \emph{normalisation}, i.e.\ the starting point of the integrals defining the formal objects~\eqref{eq:WKB_solutions}. 

To ensure that when crossing a Stokes line the associated Stokes constants are perturbative in $1/\q$, we normalise the formal solutions at the turning points from which the corresponding Stokes lines (the lines for which we evaluate $A$ and $B$) emanate. Let $u^*$ be a turning point such that $V_0(u) \sim (u-u^*)^n$ with $n \geq -1$ and $n \in \mathbb{Z}$ then \begin{equation} \label{eq:leading_ord_int}
    \int_{u^*}^uS_\text{odd, -1}(u')\, du'
\end{equation}
is well defined. However, one expects $S_{\text{odd}, m}$ to be singular at $u^*$. This can be seen by inspecting equations \eqref{eq:riccati_perturb}. This means we cannot choose integration starting from $u^*$ as the normalisation for later terms $S_{\text{odd}, j}$. Instead, there are two canonical choices of normalisation depending on whether $n$ is odd or even.

If $n$ is odd, then $S_{\text{odd},-1}$ has branch cuts that are inherited by $S_{\text{odd},j}$ and we define the formal solutions normalised at the turning point $u^*$ by 
 \begin{equation} \label{eq:WKB_sol_tp_norm_1}
    \phi_{\pm,u^*}(u) = \frac{1}{\sqrt{S_{\text{odd}}(u)}}\exp\left(\pm\frac{1}{2}\int_{\gamma_{u^*, u}} S_{\text{odd}}(u')\, du'\right)\,,
 \end{equation}
 where $\gamma_{u^*, u}$ is the path starting from $u$ in one sheet, encircling $u^*$ passing through the branch cut and ending at $u$ on the second sheet. Notice that for the leading term $S_\text{odd,-1}$ integration along  $\gamma_{u^*, u}$ reduces to the integration \eqref{eq:leading_ord_int}.
\begin{figure}[!h]
    \centering
   \begin{tikzpicture}[scale =3]
   \draw[black,thick,domain=0:180,smooth,variable=\t,
        postaction={decorate},
        decoration={markings, mark=at position 0.8 with {\arrow{<}}}]
    plot ({1.2*cos(\t)+0.5},{0.2*sin(\t)});
      \draw[black,thick,dashed, domain=180:360,smooth,variable=\t,
        postaction={decorate},
        decoration={markings, mark=at position 0.8 with {\arrow{<}}}]
    plot ({1.2*cos(\t)+0.5},{0.2*sin(\t)});
    
    \node[below] at (0.5,0.4) {$\gamma_{u^*,u}$};
    
    \filldraw[black] (1.7, 0) circle (0.8pt) node[black,above right] {$u$};
    
    \draw[decorate,blue,decoration={zigzag,segment length = 2mm,amplitude=0.5mm}] (1, 0)--(-1, 0);
       
       \filldraw[red] (1, 0) circle (1pt) node[black,above right] {$u^*$};
   \end{tikzpicture}
    \caption{The path $\gamma_{u^*,u}$ defining our normalisation around a turning point $u^*$. We draw with a solid line paths on the first sheet and with a dashed line paths on the second sheet. The zig-zagged blue line represents the branch cut.}
    \label{fig:WKB normalised at turning point}
\end{figure}
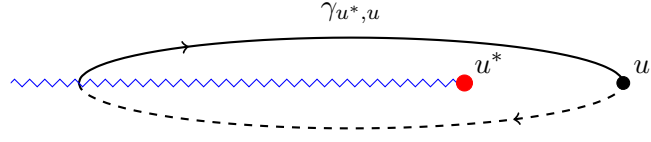

 If $n$ is even, then $S_{\text{odd},j}$ are still singular at $u^*$ but there is no branch cut structure we can exploit. Instead, we normalise our solutions in the following way
  \begin{equation} \label{eq:even_norm}
     \phi_{\pm,u^*, u_s} (u) =  \frac{1}{\sqrt{S_{\text{odd}}(u)}}\exp\left(\pm \q\int_{u^*}^u S_{\text{odd},-1}(u')\, du' \pm\int_{u_s}^u  S_{\text{odd, reg}}(u')\, du' \right)\,.
 \end{equation}
 where $u_s$ is a double pole of $V_0(u)$ and 
  \begin{equation}
     \ S_{\text{odd, reg}}(u') = S_{\text{odd}}(u')-\q \,S_{\text{odd},-1}(u')\,;\quad S_{\text{odd},-1}(u')=\sqrt{V_0(u')}\,.
 \end{equation}
 We emphasise here that these normalisations are choices. One can normalise the WKB solutions in any manner they wish and will obtain the same answers. 

 We call $u^*$ such that, $V_0(u^*) =0$, $V_0'(u^*)\neq 0$ and $u^*$ is not a singularity of $V_2(u)$ a \emph{simple turning point}. If $\phi_{\pm,u^*}(u)$ are normalised at a simple turning point then the connection formulae \eqref{eq:stokes_jump} simplify and one has $A= B =\rmi$ \cite{AIHPA_1983__39_3_211_0, Iwaki:2014vad}.

The way in which we derive connection formulae using the exact WKB method can now be stated more clearly. We start with a basis of formal solutions normalised at some point ($u =0$ in our case) defined near to some other point (also $u=0$). We then analytically continue this basis in a direction relevant for the connection problem (along the real axis to $u=1$). As this basis is analytically continued it will cross Stokes lines. If the basis is normalised at the point from which the Stokes line emanates then we cross it using the Stokes jump formulae \eqref{eq:stokes_jump}. If the basis is not normalised with respect to this Stokes line, then we must change the normalisation of the basis. For example, two formal solutions~\eqref{eq:WKB_solutions} that are normalised solely at turning points $u_1^*,\, u_2^*$ (i.e. the exponent of the leading term of the expansion of $V(u)$ at $u_1^*$ and $u_2^*$ is odd) will differ by 
\begin{equation}
    \phi_{\pm,u_1^*}(u)= \exp\left(\pm\frac{1}{2}\oint^{u_2^*}_{u_1^*}S_{\text{odd}}(u')du'\right)\phi_{\pm,u_2^*}(u) \,.
\end{equation}
If one or both of these solutions are normalised at a turning point and a pole then clearly the relevant change of normalisation will be different. The key point is that these factors are non-perturbative in $1/\q$. This is why our choices of normalisation are useful, they ensure that non-perturbative contributions come only from changing normalisation. Once the normalisation is changed, our formal solutions can cross the Stokes line and we can then change the normalisation back. Tracking all Stokes lines crossed and the corresponding discrete jumps that occur between the two points that we are connecting yields the connection formulae. 

These connection formulae can be trusted, provided that the formal WKB expansions do not break down at any of the points we are trying to connect. We discuss the appropriate method for handling these situations in the next subsection.

We conclude this subsection by discussing a useful method for determining whether a Stokes line is positive or negative when it terminates at a double pole of $V_0(u)$. Consider the quantity 
\begin{equation}
    \text{Re} \left(\q\int_{u^*}^uS_\text{odd, -1}(u')\, du' \right)
\end{equation}
with $u$ lying along the Stokes line emanating from $u^*$. If the Stokes line terminates at a double pole $u_s$ of $V_0(u)$ then near $u_s$, we have
\begin{equation} \label{eq:re_part_log}
\begin{aligned}
     \text{Re} \left(\q\int_{u^*}^uS_\text{odd, -1}(u')\, du'\right) &= \text{Re}\left(\q(\rho \log(u-u_s) +\mathcal{O}(\q^0))\right)\\
     &= \text{Re}(\q\rho) \log|u-u_s|- \text{Im}(\q\rho) \arg(u-u_s) + \mathcal{O}(1),
\end{aligned}
\end{equation}
where
\begin{equation}
    \rho  = \text{Res}_{u =u_s} S_{\text{odd,-1}}(u)
\end{equation}
is the residue of the leading term of $S_\text{odd}$. When $u$ is close to $u_s$, the sign of the quantity \eqref{eq:re_part_log} is dominated by the sign of $\text{Re}(\q\rho) \log|u-u_s|$. Specifically, \eqref{eq:re_part_log} is positive if $\text{Re}(\q\rho)<0$ and negative if $\text{Re}(\q\rho)>0$. The sign of \eqref{eq:re_part_log} does not change along a Stokes line and determines whether a Stokes line is positive or negative. Therefore, we conclude that a Stokes line ending at $u_s$ is positive if  $\text{Re}(\q\rho)<0$ and negative if $\text{Re}(\q\rho)>0$.

For the problem at hand, the leading $\q$ correction to the light-cone scaling of the shear mode frequency can be obtained with semi-classical WKB tools in a large overtone limit. However, to go beyond the first large $\q$ correction one generally needs more sophisticated methods such as a combination of the tools presented above and a way of continuing the WKB solutions to points in the complex $u$-plane   where the WKB expansions become ill-defined. This can be done effectively via the so-called uniform WKB expansions, which we now turn to.

\subsection{Uniform WKB} \label{sec:rev-uniform-WKB}

The WKB expansion is an effective tool for describing functions globally in the complex $u$-plane. However, there are specific 
values of $u$ where the WKB expansion ceases to work effectively. As mentioned briefly above, this occurs at singularities of $S_\text{odd}$ that are typically located at turning points (zeros of the leading term of the potential). Near these points, the coefficients of $S_{\text{odd}}$ become large and compete with the large-$\q$ expansion. This poses an issue if our boundary conditions are to be imposed at these problematic points.

The solution to this 
problem is the \emph{uniform WKB method} \cite{e22a9a57-b6b7-33c4-ac51-655992a56ecf, 10.1093/oso/9780198505730.001.0001}. 
The idea is to use an approximate form of the differential equation in the vicinity of a turning point so as to obtain an approximation which is uniform as the independent variable $u$ varies over the vicinity of a turning point. The WKB ansatz is replaced by the form 
\begin{equation}
\label{liouville}
    \phi(u) = \frac{1}{\sqrt{\xi'(u)}}G(\xi(u))
\end{equation}
where $\xi(u)$ and $G(\xi)$ will be determined appropriately. $\xi(u)$ is called the \emph{generalised coordinate}. Transformations of this form have the property that they map the original differential equation to one of the same (normal) form:  
\begin{equation} \label{eq:local_ODE}
    G''(\xi)-V_L(\xi) G(\xi) =0.
\end{equation}
with the new potential $V_L$ related to the function $\xi$ by
\begin{equation} \label{eq:gen_coord_eqn}
    \frac{1}{2}\{\xi(u), u\} - (\xi'(u))^2V_L(\xi(u)) +V(u)=0.
\end{equation}
where 
\begin{equation}
     \{\xi(u), u\}  =\frac{\xi'''(u)}{\xi'(u)}-\frac{3}{2}\left(\frac{\xi''(u)}{\xi'(u)}\right)^2.
\end{equation}
The new potential, $V_L(\xi)$, is called the \emph{local potential} and $\xi(u)$ is called the \emph{generalised coordinate}. If we want a uniform description of our function for the typical WKB range of validity, large-$\q$, but also at a turning point, $
u=u^*$, then the potential $V_L(\xi)$ is chosen such that it captures the behaviour of $V(u)$ near $u=u^*$. Once this choice is made, the coordinate function $\xi(u)$ can be found by solving the differential equation \eqref{eq:gen_coord_eqn}. If no such $\xi(u)$ obeying $\xi(u^*)=0$ can be found, then one may need to modify $V_L(\xi)$ by including subleading terms that match the expansion of $V(u)$.

In general, it is unlikely that one can solve \eqref{eq:gen_coord_eqn} exactly, so instead  $\xi(u)$ is found via a large-$\q$ expansion:
\begin{equation}
\label{eq:xi-expansion-q}
    \xi(u) = \sum_{n=0}^\infty\,
\q^{-n}\xi_n(u)\end{equation}
where the powers $\q$ appearing in this expansion can be deduced from equation \eqref{eq:gen_coord_eqn}. Each coefficient $\xi_n(u)$ will obey a first order differential equation and will require boundary conditions to fully specify. We select the boundary condition $\xi_n(u^*)=0$ so that the behaviour of $\phi(u)$ near $u=u^*$ is determined by the behaviour of $G(\xi)$ near $u=u^*$. The first order differential equations are non-linear and it quickly becomes impractical to solve for each $\xi_n(u)$ exactly. However, if we are interested in matching the Frobenius expansion at $u =u^*$ to the large-$\q$ WKB expansions then we require only the local expansions of $\xi_n(u)$ near to $u =u^*$:
\begin{equation}
\label{eq:xi-expansion-u}
\xi_n(u)=\sum_{m=1}^\infty c_{n,m}(u-u^*)^m.
\end{equation}
This procedure generalises the well-known treatment of $1$-dimensional Schroedinger eigenvalue problems. Within the WKB approach, where a solution is sought in the limit $\hbar\to0$, near a simple turning point the potential energy is approximated by a linear potential, which leads to the Airy equation. The process outlined above is formulated in a way which is applicable to more intricate situations, such as the one encountered further on in this paper, where the expansion parameter $\q$ enters in a way which affects the behaviour of the potential near turning points. 

\section{Large-\texorpdfstring{$\q$}{q} corrections to imaginary light-cone scaling} 
\label{sec:lc-1}

In Section~\ref{sec:lc-scaling} we argued that in imaginary light cone scaling 
\begin{equation}
\label{eq:qeta}
   \w\sim\q\quad \mathrm{with} \quad \q=\rmi\eta,\quad \eta\to+\infty \,,
\end{equation} 
we had $\w \sim \q$ as a possible solution to the QNM problem. Writing $\w = s \,\q$ as before, assume that
\begin{equation}
\label{eq:sdef}
    s(\eta)^2 = 1-\eta^{-\lambda}\kappa(\eta),
\end{equation}
where $\lambda>0$ and 
\begin{equation}
\label{eq:kappa-eta}
    \kappa(\eta) = k_{0}+\dots \,,
\end{equation}
with $k_{0}>0$ as $0<s(\eta)^2\leq1$ for positive $\eta$. The ellipses denote higher order terms as a power series in negative powers of $\eta$ that each go to zero as $\eta \to \infty$.  We also found, by a scaling argument for the bound states of eq.~\eqref{eq:Schreq-Phi}, that $\lambda = 2(p+1)/(p+3)$, and that $k_0>0$ is fixed by the ground-state energy of the boundary-layer problem~\eqref{eq:Schreq-BL}, consistent with causality.

The purpose of this section is to recover this result within the exact WKB framework that we use in the rest of the paper.  This is not just a re-derivation of the results of Section~\ref{sec:lc-scaling}. First, the argument of Section~\ref{sec:lc-scaling} relies on the problem being self-adjoint (which requires $\q$ to be exactly imaginary). The WKB method used below makes no such assumption of self-adjointness. 

It allows us to further check the (non-)existence of non-perturbative corrections to the frequency in the large-$\q$ expansion. Finally, it isolates the inner-outer structure of the problem in the original coordinate $u$, avoiding the tortoise coordinate~$y$. It is this structure that we extend to large overtones in Section~\ref{sec:large_n}, and to all orders in $1/\eta$ for $p=3$ in Section~\ref{sec:all_orders}, where the expansion~\eqref{eq:kappa-eta} is determined to arbitrary order. We present only the leading large overtone result in Section~\ref{sec:large_n}, but it is straightforward to extract higher perturbative terms in inverse powers of the overtone $n$.

\subsection{Connection problem at leading order} 
\label{sec:rev-leading-all-p}

To derive the quantisation condition obeyed by $\kappa(\eta)$ we need to connect the Frobenius expansions at $u=0$ to the Frobenius expansions at $u=1$. We do this by constructing large-$\eta$ WKB solutions and keeping track of Stokes phenomena as these WKB solutions are analytically continued from $u=1$ to $u=0$. There is a turning point at $u =\infty$ whose associated Stokes jumps may be involved in the connection formulae. To ensure that no Stokes lines emanating from $u =\infty$ are missed, we change the coordinate of \eqref{eq:Srchodinger-eq-u} from $u$ to 
\begin{equation}
    u = \frac{\zeta}{2-\zeta}.
\end{equation}
This coordinate change fixes $0$ and $1$ while $u=\infty$ is mapped to $\zeta =2$. It also introduces a first derivative that we remove by setting
\begin{equation}
    \phi(\zeta) = (2-\zeta)^{-1}\chi(\zeta).
\end{equation}
The equation for the shear mode \eqref{eq:Srchodinger-eq-u} in terms of $\chi(\zeta)$ and $\zeta$ is
\begin{multline}\label{eq:zeta_eqn}
      \chi''(\zeta) -\left[\eta^2\frac{(p+1)^2\zeta^{1+p}(2-\zeta)^{p-3}}{\left((2-\zeta)^{1+p}-\zeta^{1+p}\right)^2}-\eta ^{2-\lambda}\frac{(1+p)^2  (2-\zeta)^{2p-2}\kappa(\eta)}{\left((2-\zeta)^{1+p}-\zeta^{1+p}\right)^{2}}\right. \\ \left. +  \frac{(p-2)p(2-\zeta)^{2+2p}-(1-4p^2)\zeta^{2+2p}-6p^2((2-\zeta)\zeta)^{1+p}}{\zeta^2(2-\zeta)^2\left((2-\zeta)^{1+p}-\zeta^{1+p}\right)^2}\right]\chi(\zeta) =0
\end{multline}
and the boundary conditions obeyed by the QNM frequency are
\begin{equation}\label{eq:chi_bc}
    \chi(\zeta) = \begin{cases}
        \zeta^{p/2}, \quad &\zeta \to 0,\\
        (1-\zeta)^{1/2 -\rmi \w /2} \quad &\zeta\to 1.
    \end{cases}
\end{equation}
Recall that we are taking $\kappa(\eta)$ as being of leading order $\mathcal{O}(\eta^0)$ at large $\q$.

\begin{figure}[htbp] \captionsetup[subfigure]{labelformat=empty}\centering  \begin{subfigure}{0.32\textwidth} \centering \includegraphics[width=\linewidth]{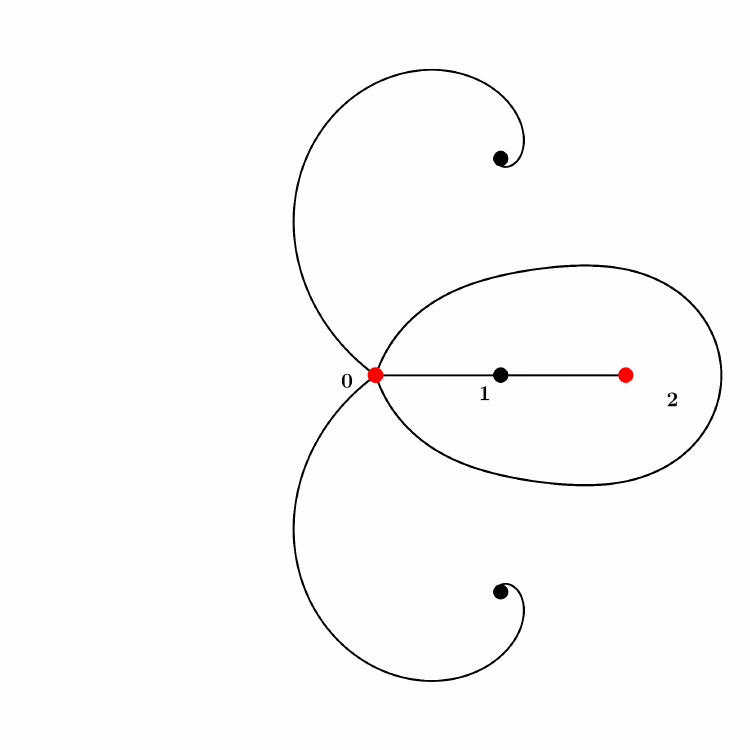} \caption{$p=2$}\end{subfigure} 
\begin{subfigure}{0.32\textwidth} \centering \includegraphics[width=\linewidth]{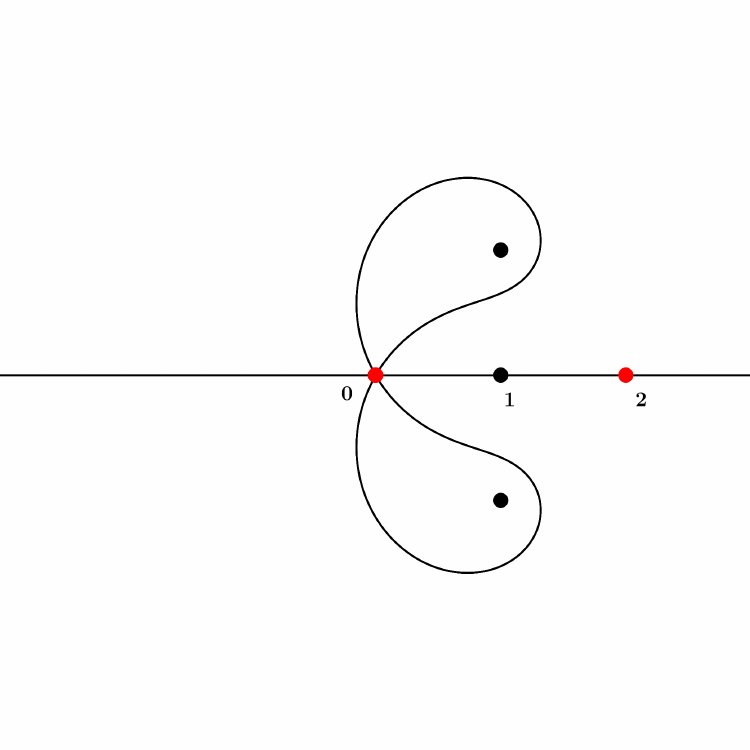} \caption{$p=3$}  \end{subfigure} 
\begin{subfigure}{0.32\textwidth} \centering \includegraphics[width=\linewidth]{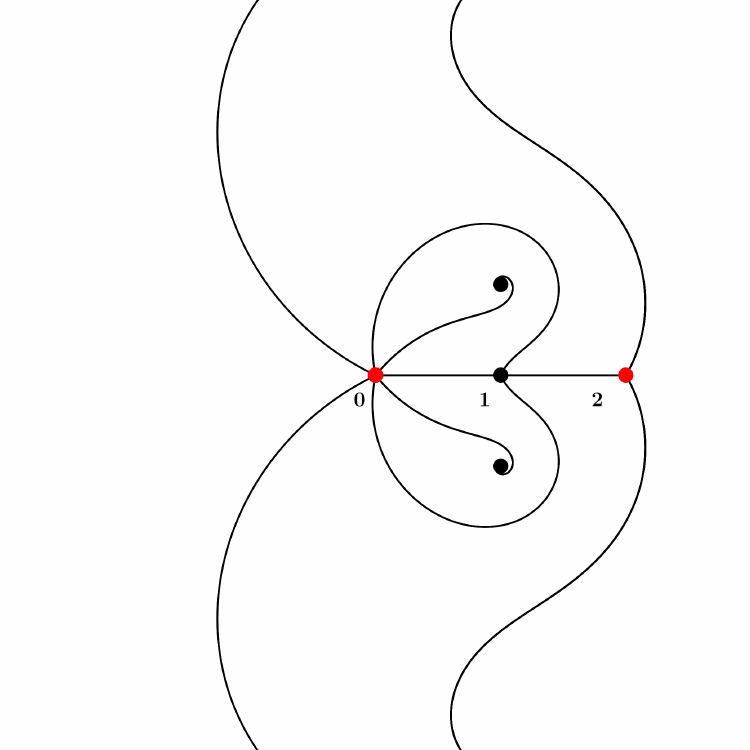} \caption{$p=4$}  \end{subfigure} \vspace{0.5cm}
\centering  \begin{subfigure}{0.32\textwidth} \centering \includegraphics[width=\linewidth]{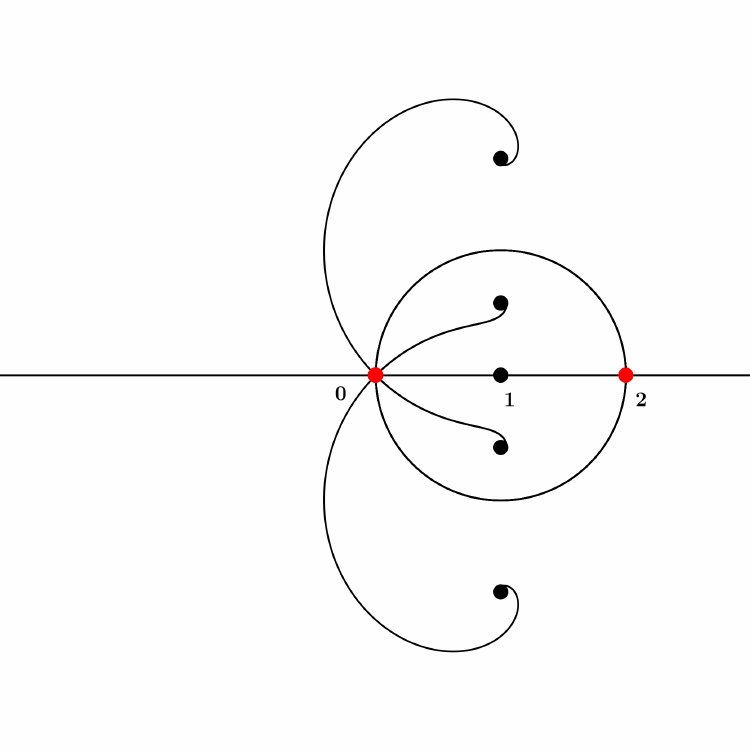} \caption{$p=5$}\end{subfigure} 
\begin{subfigure}{0.32\textwidth} \centering \includegraphics[width=\linewidth]{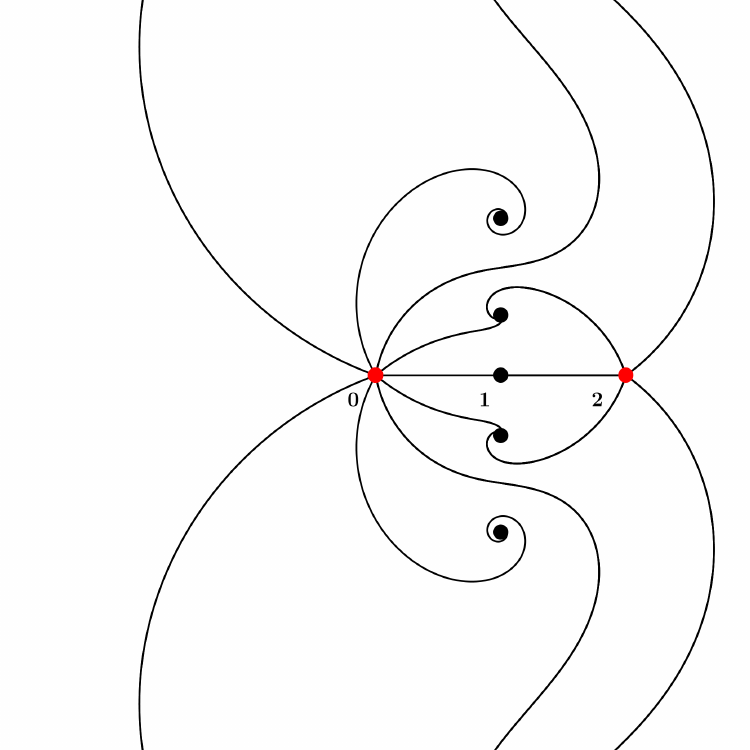} \caption{$p=6$}  \end{subfigure} 
\begin{subfigure}{0.32\textwidth} \centering \includegraphics[width=\linewidth]{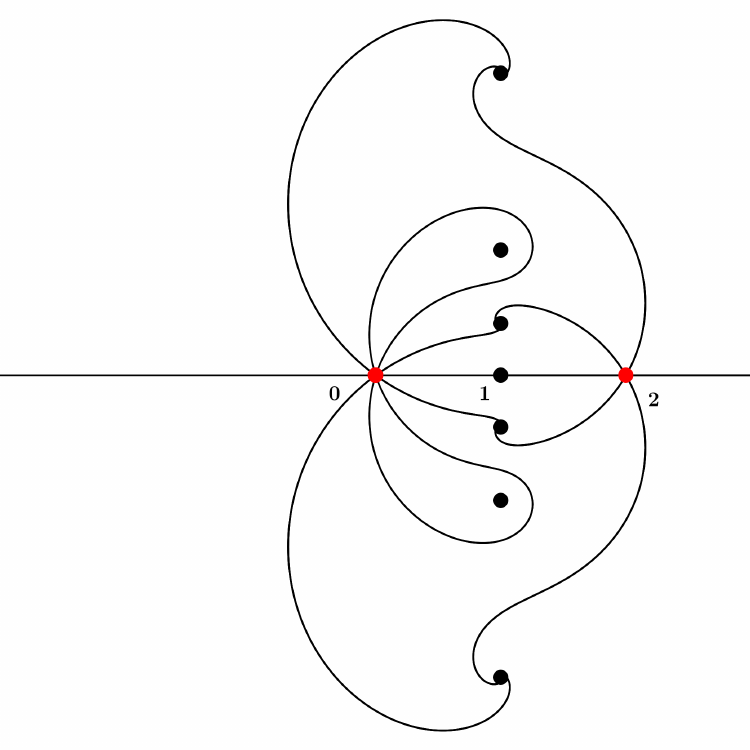} \caption{$p=7$}  \end{subfigure} \vspace{0.5cm}

\begin{subfigure}{0.32\textwidth} \centering \includegraphics[width=\linewidth]{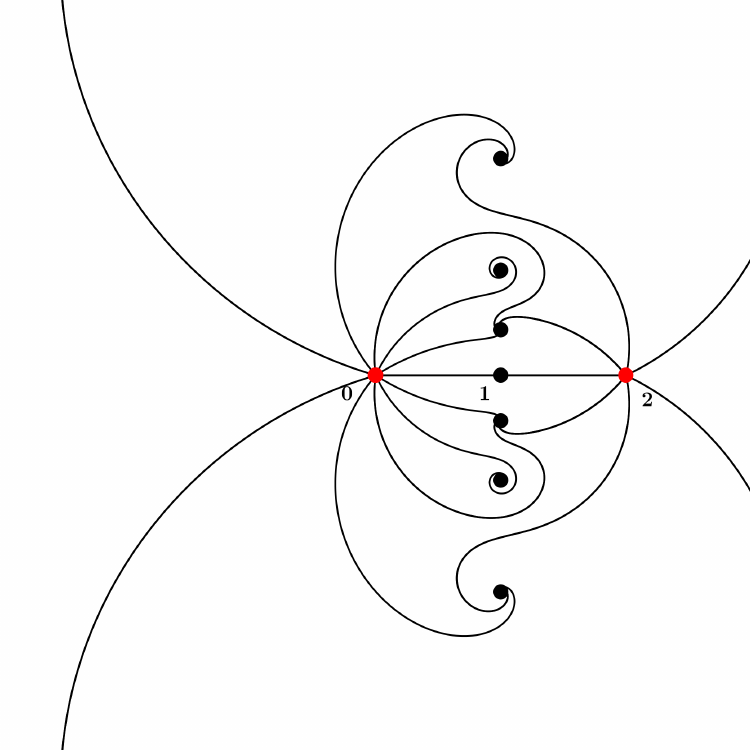} \caption{$p=8$}  \end{subfigure} 
\begin{subfigure}{0.32\textwidth} \centering \includegraphics[width=\linewidth]{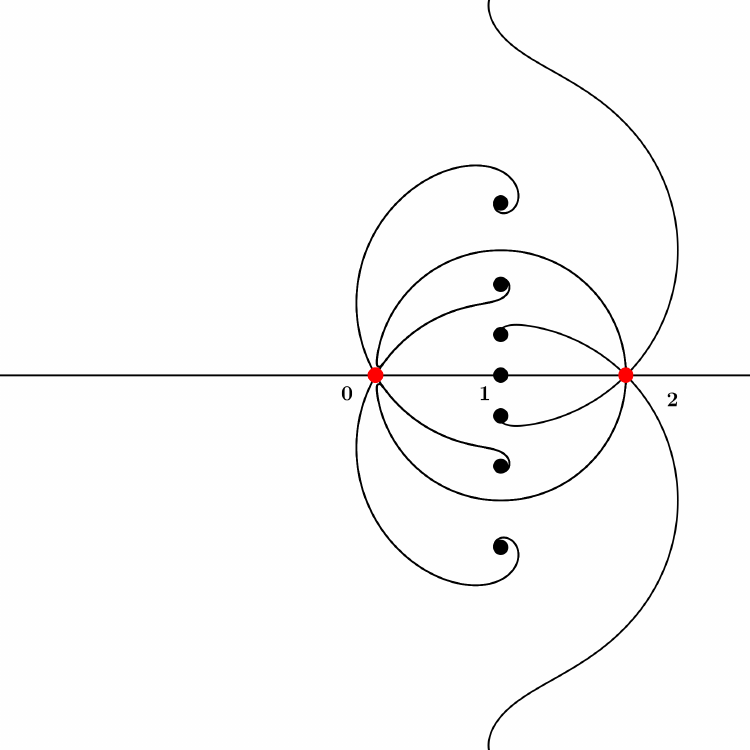} \caption{$p=9$}  \end{subfigure} \vspace{0.5cm}
\caption{Stokes graphs associated to equation \eqref{eq:zeta_eqn}, in the complex $\zeta$ plane, for $\arg\q =\pi/2$ and $p =2, \dots, 9$. The black dots are the double poles of the leading term in the potential, \eqref{eq:zeta_eqn} and the red dots are the turning points located at $\zeta =0, 2$. We have not shown the branch cuts in these plots.
} 
\label{fig:SG1} \end{figure}

\paragraph{Connection formula} 
The Stokes graphs associated to equation \eqref{eq:zeta_eqn} are governed by 
\begin{equation}
    S_{\text{odd},-1}(\zeta) =-(p+1)\frac{\zeta^{(1+p)/2}(2-\zeta)^{(p-3)/2}}{\left(\zeta^{1+p}-(2-\zeta)^{1+p}\right)}
\end{equation}
The Stokes graphs have Stokes lines emanating from the turning points at $\zeta=0$ and $\zeta=2$; they are shown in Figure \ref{fig:SG1} for various dimensions and for $\arg(\eta) = 0$.

In all Stokes graphs, there is a Stokes line that lies on the physical domain $\zeta\in [0,1]$. As discussed at the end of subsection \ref{sec:rev-WKB}, whether a Stokes line is positive or negative can be determined by the real part of the residue of $S_{\text{odd},-1}(\zeta)$ at the point where the Stokes line ends. We find
\begin{equation}
   \text{Res}_{\zeta =1}\eta S_\text{odd,-1}(\zeta) = -\frac{\eta}{2}.
\end{equation}
Thus, if $\pi/2>\arg(\eta)>-\pi/2$, the real part of this quantity is negative and any Stokes line ending at $\zeta =1$ is a positive Stokes line. This fact is crucial in simplifying our analysis as we will now explain.

First, consider the linear combination of WKB solutions required to satisfy the boundary condition at $\zeta =1$. At leading order in $\eta$, the large-$\eta$ WKB solutions $\chi_\pm(\zeta)$ are given by 
\begin{equation}\label{eq:leading_large_q}
    \chi_\pm (\zeta) = \frac{1}{\sqrt{\eta S_{\text{odd,$-1$}}(\zeta)}}\exp \left( \pm \eta\int^\zeta S_{\text{odd,-1}}(\zeta')\, d\zeta'\right)\left(1 + \mathcal{O}\left(\eta^{1-\lambda}\right)\right).
\end{equation}
 Expanding this for $\zeta$ close to $1$, we find
\begin{equation}
       \chi_\pm (\zeta) \propto (\zeta-1)^{1/2 \mp\eta/2}(1 +\mathcal{O}(\zeta -1)) + \dots .
\end{equation}
Comparing this expansion with the boundary condition at $\zeta=1$, \eqref{eq:chi_bc}, we see that only $\chi_-(\zeta)$ can satisfy this boundary condition. Now consider starting with $\chi_-(\zeta)$ close to $\zeta =1$ and analytically continue it to small $\zeta$ close to $0$, but not all the way to $\zeta=0$. This solution lies along a positive Stokes line  and will not have any Stokes jumps. Therefore, our boundary condition at $\zeta=1$ is satisfied if the large-$\eta$ expansion of our solution is $\chi_-(\zeta)$ close to $\zeta =0$. 

The boundary value problem defining the shear mode is solved if $\chi(\zeta) \sim \zeta^{p/2}$ as $\zeta \to 0$ and $\chi(\zeta) \sim \chi_-(\zeta)$ for large-$\eta$ and positive $\zeta$ close to $\zeta =0$. These requirements do not depend on the Stokes graph and certainly do not depend on Stokes lines emanating from $\zeta=2$ or $u =\infty$. For this reason, we now work with the original coordinate $u$.

\paragraph{Inner-Outer matching}
Now let us match the Frobenius expansion about $\zeta=0$ to $\chi_-(\zeta)$, which in the $u$-coordinate becomes the matching of the Frobenius expansion at $u=0$ to $\phi_- (u)$. Our goal is to find what linear combination of large-$\eta$ WKB solutions behave as $u^{p/2}$ as $u \to 0$. If we attempt to expand the leading large-$\eta$ term of $\phi_\pm(u)$ close to $u=0$, we find 
\begin{equation}
      \phi_\pm (u) \propto\frac{1}{\sqrt{\eta u^{(1+p)/2}}}\exp \left( \pm \eta \frac{1+p}{3+p}u^{(3+p)/2}\right)(1+\dots).
\end{equation}
As $u\to 0$, the factor $u^{(1+p)/2}$ becomes small and competes with the large-$\eta$ expansion. Explicitly, when $ \eta^{2/(p+3)}u \sim 1$ the argument of the exponential is $\mathcal{O}(\eta^0)$ and we expect that all terms of the WKB solutions are of the same order. Therefore, we cannot trust the small-$u$ expansion of the WKB solutions. 

To find what linear combination of WKB solutions yields the correct small-$u$ behaviour we could use the uniform WKB method described in section \ref{sec:rev-uniform-WKB}. However, this method is overkill for the current discussion. Instead, we find the appropriate linear combination by rescaling to the coordinate 

\begin{equation} \label{eq:leading_order_rescale}
     \tau = \eta^{2/(3+p)} u.
 \end{equation}
 Small $\tau$ corresponds to small $u$ but large real and positive $\tau$ corresponds to large positive $\eta$. The large-$\tau$ regime overlaps with the large-$\eta$. If we can find the linear combination of large-$\tau$ solutions that provide the correct behaviour at small $\tau$, then the linear combination of WKB solutions is found by comparing the two expansions in the overlapping region.

At leading orders in $\eta$, equation \eqref{eq:Srchodinger-eq-u} in terms of $\tau$ is
\begin{multline} \label{eq:tau_eqn1}
      \phi''(\tau) -\left[\frac{(p+1)^2}{4}\tau^{p+1}+\frac{p(p-2)}{4 \tau^2}\right.\\ - \left.\eta^{2(p+1)/(p+3)-\lambda}\frac{(p+1)^2}{4} k_0 + \mathcal{O}\left(\eta^{-2(p+1)/(p+3)}\right)\right]\phi(\tau) =0.
\end{multline}

If we assume that $\lambda>2(p+1)/(p+3)$, then the leading order solution $\phi(\tau)$ that obeys the boundary conditions at $u = 0$ is 
\begin{equation}\label{eq:leading_ord_soln_wrong}
     \phi(\tau)\propto \sqrt{\tau}I_{\frac{p-1}{p+3}}\left( \frac{p+1}{p+3}\tau^{(p+3)/2}\right) + h.o.t.
\end{equation}
The large-$\tau$ expansion of the above expression always has an exponentially enhanced and an exponentially suppressed term regardless of $k_0$, see \cite[\href{https://dlmf.nist.gov/10.40}{(10.40.5)}]{NIST:DLMF}. However, for the boundary condition at $u=1$ to be satisfied, the large-$\tau$ expansion must contain only an exponentially suppressed term that is proportional to $\phi_-(u)$. Therefore, the two solutions cannot be matched and we must have
\begin{equation}
   \lambda \leq \frac{2(p+1)}{(p+3)}.
\end{equation}
 This implies that $\lambda \leq 2(p+1)/(p+3)$. If we assume that $\lambda < 2(p+1)/(p+3)$ then the inner-outer matching near $u=0$ is more involved. There are now three small-$u$ regions that need to be matched: the large-$\eta$ WKB regime must be matched to a regime where $u \sim \eta^{-\lambda/(p+1)}$ which then has to be matched to the Frobenius expansion at $u=0$. We would find that there are no such solutions in this case meaning that
\begin{equation} \label{eq:lambda_scaling}
    \lambda= \frac{2(p+1)}{(p+3)}.
\end{equation}
Ultimately, we could treat the scaling \eqref{eq:lambda_scaling} as an ansatz that leads to a consistent answer for $\kappa(\eta)$, which agrees with Section~\ref{sec:lc-scaling}.

This scaling introduces the $k_0$ dependent term in the leading order potential of the rescaled equation \eqref{eq:tau_eqn1}. Enforcing the boundary conditions at $\tau =0$ and $\tau=\infty$ would then quantise $k_{0}$. We discuss an effective method for calculating many terms of the large-$\eta$ expansion of $\kappa(\eta)$ for $p=3$ in the next section. However, before this, we derive a large overtone formula for $k_{0}$.

\subsection{Large-\texorpdfstring{$n$}{n} formula}\label{sec:large_n}

To calculate $k_{0}$, we solve the leading order large-$\eta$ equation \eqref{eq:tau_eqn1} with $\lambda = 2(p+1)/(p+3)$
\begin{equation}
    \phi''(\tau) -\left[\frac{(p+1)^2}{4}\tau^{p+1}+\frac{p(p-2)}{4 \tau^2} -\frac{(p+1)^2}{4} k_{0}\right]\phi(\tau) =0.
\end{equation}
subject to the boundary conditions
\begin{equation}\label{eq:LC_boundary_conditions}
    \phi(\tau) \sim \begin{cases}\tau^{p/2}, \quad &\tau\to 0,\\
    \tau^{-(p+1)/4}\exp\left[- \frac{(p+1)}{(p+3)}\tau^{(p+3)/2}\right], &\tau\to +\infty.
    \end{cases}
\end{equation}
The first boundary condition corresponds to the original boundary requirement at $u=0$ \eqref{eq:bc_u} and the second ensures that the solution is proportional to the large-$\eta$ WKB solution $\phi_-(u)$. 

\begin{figure}[htbp] \captionsetup[subfigure]{labelformat=empty}\centering  \begin{subfigure}{0.32\textwidth} \centering \includegraphics[width=\linewidth]{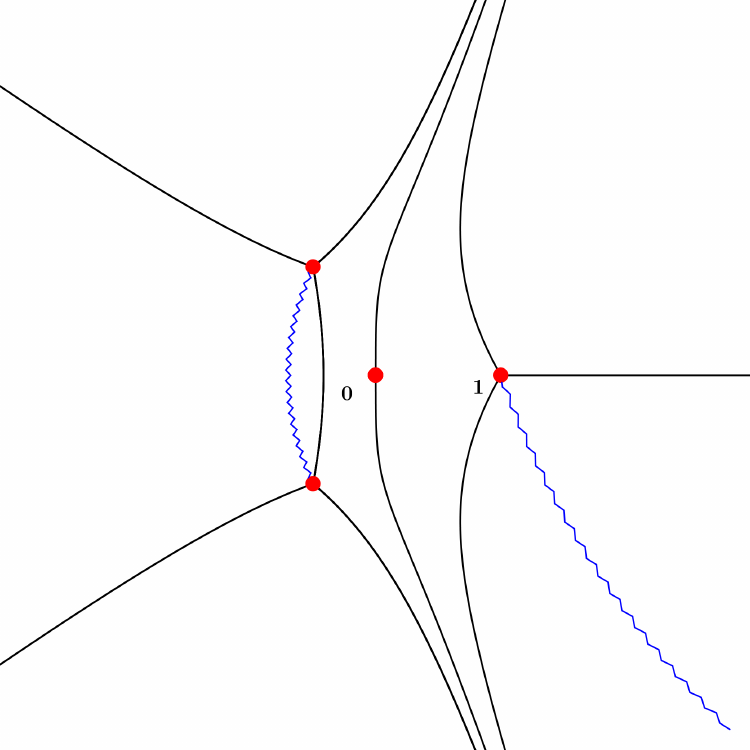} \caption{$p=2$}\end{subfigure} 
\begin{subfigure}{0.32\textwidth} \centering \includegraphics[width=\linewidth]{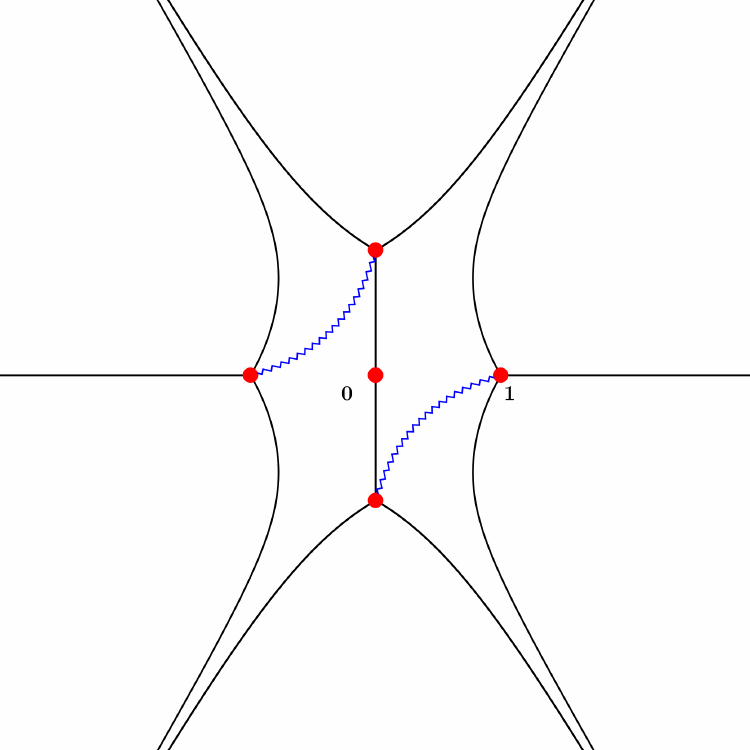} \caption{$p=3$}  \end{subfigure} 
\begin{subfigure}{0.32\textwidth} \centering \includegraphics[width=\linewidth]{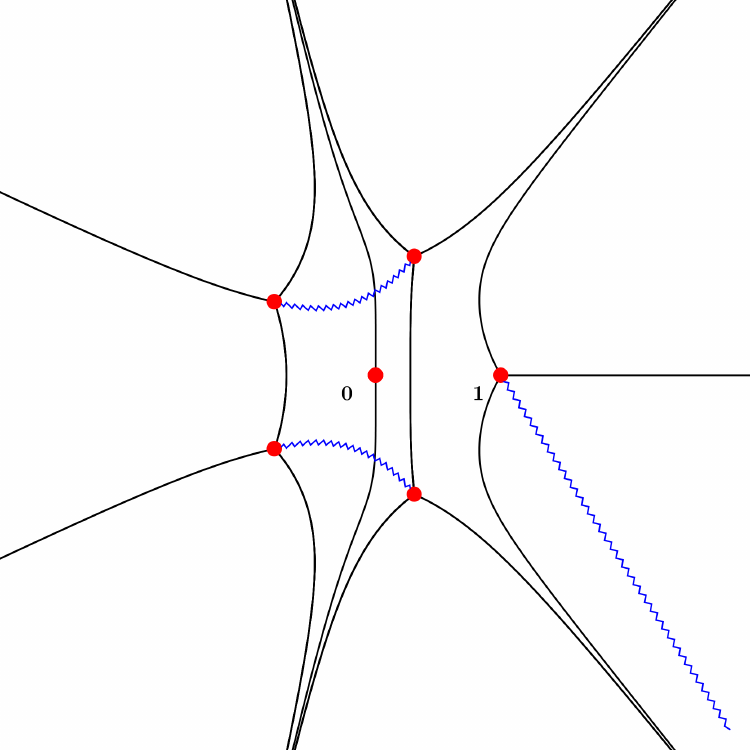} \caption{$p=4$}  \end{subfigure} \vspace{0.5cm}
\centering  \begin{subfigure}{0.32\textwidth} \centering \includegraphics[width=\linewidth]{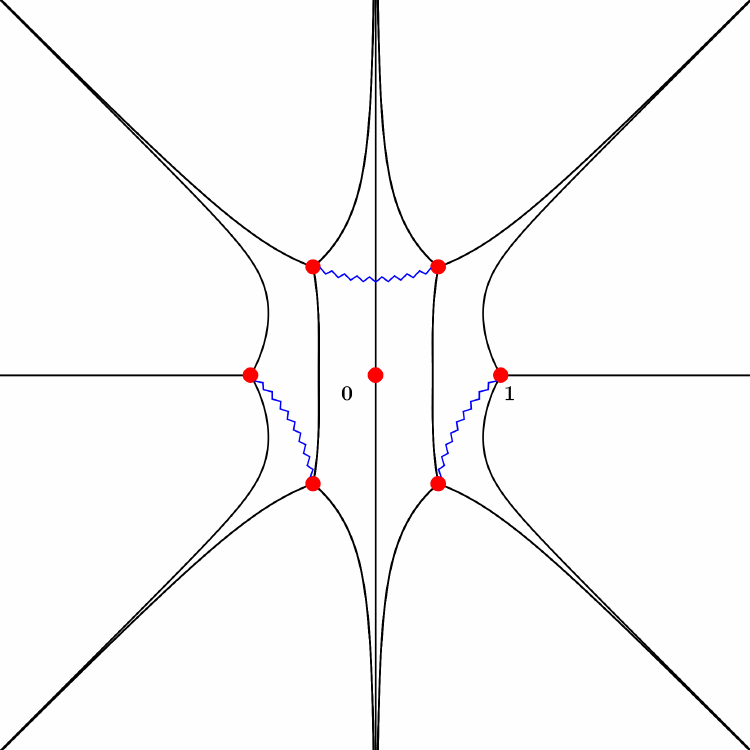} \caption{$p=5$}\end{subfigure} 
\begin{subfigure}{0.32\textwidth} \centering \includegraphics[width=\linewidth]{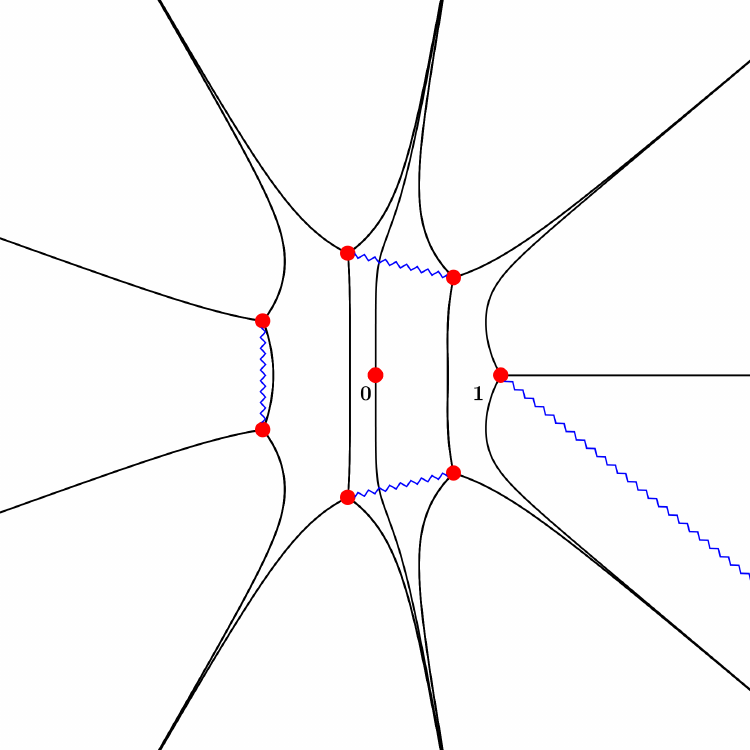} \caption{$p=6$}  \end{subfigure} 
\begin{subfigure}{0.32\textwidth} \centering \includegraphics[width=\linewidth]{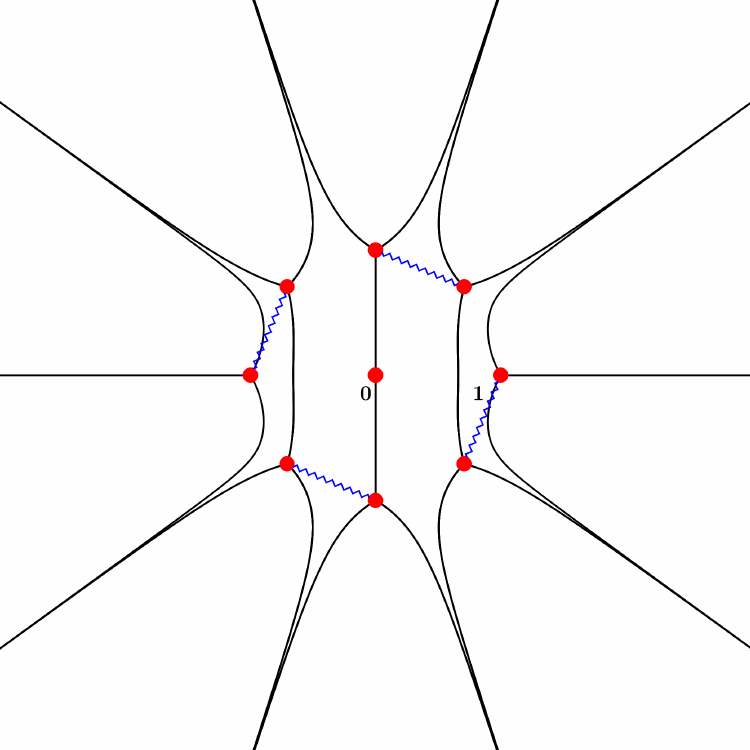} \caption{$p=7$}  \end{subfigure} \vspace{0.5cm}

\begin{subfigure}{0.32\textwidth} \centering \includegraphics[width=\linewidth]{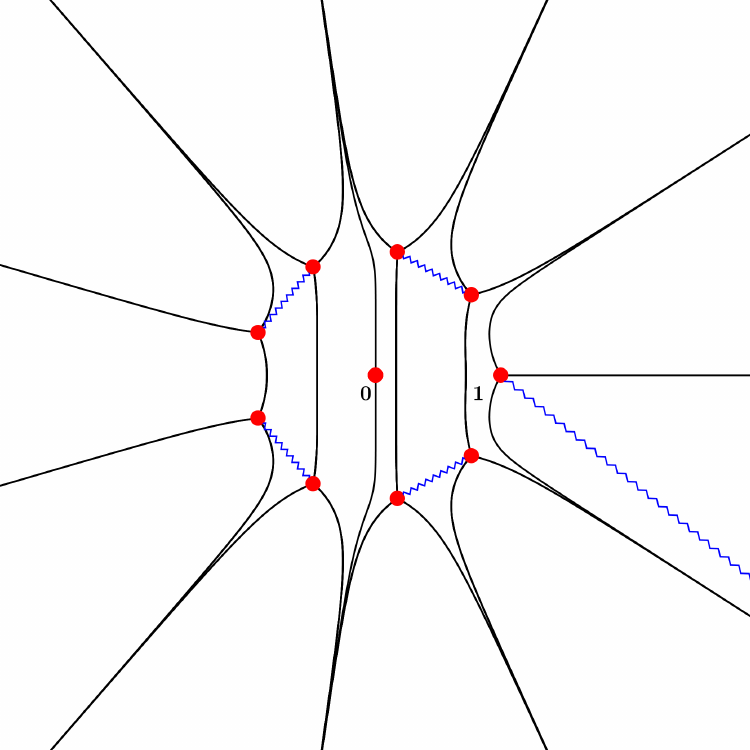} \caption{$p=8$}  \end{subfigure} 
\begin{subfigure}{0.32\textwidth} \centering \includegraphics[width=\linewidth]{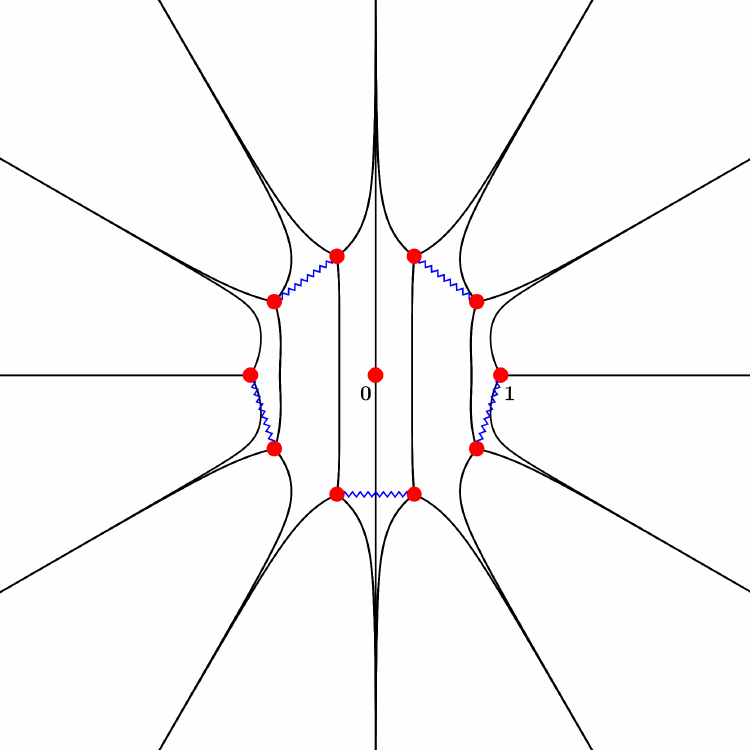} \caption{$p=9$}  \end{subfigure} \vspace{0.5cm}
\caption{Stokes graphs associated to the problem \eqref{eqn:LC_inner_leading_eqn} for $\hbar$ real and positive. The red dots are turning points, the black lines are Stokes lines and the blue zig-zagged lines are branch cuts. From $p\geq 4$ onwards, there are bounded Stokes lines connecting turning points that will contribute non-perturbative terms in $n$ that are ignored in our analysis.
} 
\label{fig:SG2} \end{figure}

The boundary value problem \eqref{eq:LC_boundary_conditions} can be solved numerically to determine $k_0$ (see \cite{Fuini:2016qsc} for a similar calculation in the scalar-channel). In the spirit of \cite{Aniceto:2026zmc}, we take an alternative route and find a large-$n$ formula for $k_{0}$, where $n$ is the overtone of the QNM frequency. This can be achieved by assuming that $k_{0}$ is large.  First, we define 
\begin{equation}
    \hbar^{-1} =\frac{(p+1)}{2}\left(k_{0}\right)^{\frac{p+3}{2(1+p)}}.
\end{equation}
Note that $\hbar >0$ because $k_{0}>0$. We then rescale with respect to $k_{0}$ by setting $\tau = k_{0}^{1/(p+1)}x$, and the resulting equation is 
\begin{equation} \label{eqn:LC_inner_leading_eqn}
    \phi''(x) -\left[\hbar^{-2}\left(x^{p+1}-1\right)+\frac{p(p-2)}{4x^2}\right]\phi(x) =0.
\end{equation}
with $\hbar$ small, real and positive. The boundary conditions \eqref{eq:LC_boundary_conditions} in terms of the new variable $x$ are 
\begin{equation}\label{eq:LC_boundary_conditions_x}
    \phi(x) \sim \begin{cases}x^{p/2}, \quad &x\to 0,\\
    x^{-(p+1)/4}\exp\left[-\frac{2}{\hbar} \frac{x^{(p+3)/2}}{(p+3)}\right], &x\to +\infty.
    \end{cases}
\end{equation}
Our goal is to calculate the large-$n$ expansion of $\hbar$. This is achieved by solving the boundary value problem via a small-$\hbar$ WKB analysis. The first step is to plot the Stokes graphs for $\arg(\hbar) =0$, these are produced by plotting the Stokes lines
\begin{equation}
    \text{Im} \left(\hbar \int_{x_p}^x \sqrt{(x')^{p+1}-1} \, dx'\right)=0
\end{equation}
where\footnote{Although $x=0$ is not a zero of the leading term of the potential, it is a singularity of $S_\text{odd}$ due to the subleading power of $x^{-2}$. Thus, we should expect Stokes phenomena associated to the point $x=0$.} $x_p=0$ or $\rme^{\rmi \pi m /(p+1)}$ with $m= 0, \dots, p$. The Stokes graphs for $p=2, \dots, 9$ are shown in Figure \ref{fig:SG2}.

The aim now is to analytically continue the WKB solutions from $x=\infty$ to $x=0$. However, as $p$ increases there are a number of lines that connect two turning points and intersect the interval $[0, +\infty)$. These Stokes lines must be crossed when analytically continuing from $x= +\infty$ to $x =0$ and will contribute to the connection formula. Stokes lines that start and end at turning points are called \emph{bounded Stokes lines} and typically control non-perturbative contributions \cite{AIHPA_1983__39_3_211_0, Delabaere1997ExactSE}. However, in our current discussion, we are only interested in a perturbative large-$n$ formula for $k_{0}$. For this reason, we will ignore the bounded Stokes lines and accept that the derived large-$n$ formula may be incomplete.

If we ignore the bounded Stokes lines, then the connection problem is very simple. We start with a basis of WKB solutions near to $x=+\infty$ normalised at the turning point located at $x =1$. The solutions $\phi_{\pm,1}(x)$ can be analytically continued close to $x =1$ without crossing any Stokes line. Once the basis of solutions $\phi_{\pm, 1}(x)$ is close to $x=1$, we can analytically continue above or below the turning point $x=1$. We choose to analytically continue above $x=1$ to avoid the branch cuts. We could have chosen to analytically continue below, this would lead to a slightly different Stokes jump and one would have to account for the WKB solutions crossing the branch cut. Ultimately, the choice will not matter, and we choose the simplest option.

When analytically continuing in this direction, we cross the Stokes line emanating from $x =1$ in the anticlockwise direction. This Stokes line is a negative Stokes line associated to a simple turning point and our WKB basis is already normalised at $x=1$. Therefore, using formula \eqref{eq:stokes_jump}, the analytic continuation when crossing this Stokes line is 
\begin{equation}\label{eq:connection_form_LC}
  \left(  \begin{matrix}
        \phi_{+,1}\\
        \phi_{-, 1}
    \end{matrix}\right) =  \left(  \begin{matrix}
       1 & 0\\
       \rmi &1
    \end{matrix}\right) \left(  \begin{matrix}
        \phi_{+,1}\\
        \phi_{-, 1}
    \end{matrix}\right).
\end{equation}
The WKB solutions can then be continued to $x =0$. In a similar fashion to Section \ref{sec:rev-leading-all-p}, we would naively find how the basis of WKB solutions is related to the Frobenius expansion at $x=0$ by expanding our WKB solutions close to $x=0$. However, as $x\to 0$ one finds that the terms in expansion of $S_\text{odd}(x)$ associated to $\phi_\pm(x)$ behave as 
\begin{equation}
    \hbar^{2n-1} S_{\text{odd},2n-1}(x)\sim \hbar^{2n-1}x^{-2n}.
\end{equation}
This implies that as $x\to 0$, there will be a point at which $x\sim \hbar$ and all terms of the small-$\hbar$ expansion are $\mathcal{O}(\hbar^0)$ causing the WKB expansion to break down. To resolve this issue, we do not use inner-outer matching as in Section \ref{sec:rev-leading-all-p} and instead use the uniform WKB method to find how Frobenius expansions at $x =0$ are related to the small-$\hbar$ WKB expansions.

To use the uniform WKB method, we need to specify a local potential $V_L(\xi)$. The expansion of  \eqref{eqn:LC_inner_leading_eqn} around $x=0$ motivates the local potential:
\begin{equation} \label{eq:bessel_Potential}
    V_L(\xi) = - \frac{1}{4\hbar^2}+\frac{p(p-2)}{4\xi^2}.
\end{equation}
The solution to the equation with potential \eqref{eq:bessel_Potential} that is regular at $\xi =0$ is 
\begin{equation}
    \sqrt{\xi} J_{(p-1)/2}\left(\frac{\xi}{2 \hbar}\right)
\end{equation}
where $J_\nu(\xi)$ is the Bessel function of first kind. By expanding this solution for small-$\hbar$ we find 
\begin{equation}
    \sqrt{\xi} J_{(p-1)/2}\left(\frac{\xi}{2 \hbar}\right) \propto \rme^{- \rmi p \pi /2}\varphi_{+, 0, \infty}(\xi) + \varphi_{-, 0, \infty}(\xi)
\end{equation}
where $\varphi_{\pm, 0, \infty}(\xi)$ are the small-$\hbar$ WKB solutions of the equation with potential \eqref{eq:bessel_Potential} normalised at the turning point $\xi=0$ and at $\xi=\infty$ as in equation \eqref{eq:even_norm}.

This implies that the solution satisfying the boundary condition at $x = 0$ can be written as 
\begin{equation}\label{eq:local_WKB_sol_bes}
\begin{aligned}
    \phi(x) &\propto \sqrt{\frac{\xi(x)}{\xi'(x)}} J_{(p-1)/2}\left( \frac{\xi(x)}{2\hbar}\right)\\
    &\propto \frac{1}{\sqrt{\xi'(x)}} \left[\rme^{- \rmi p \pi /2}\varphi_{+, 0, \infty}(\xi(x)) + \varphi_{-, 0, \infty}\left(\xi(x)\right)\right].
    \end{aligned}
\end{equation}
 $\xi(x)$ is the generalised coordinate that can be constructed as a small-$\hbar$ and small-$x$ expansion as explained in Section \ref{sec:rev-uniform-WKB}.

The solutions $\varphi_{\pm , 0, \infty}\left(\xi(x)\right)/\sqrt{\xi'(x)}$ are small-$\hbar$ expansions that are exponentially enhanced or suppressed. Therefore, these solutions must be proportional to the original WKB solutions $\phi_{\pm,1}(x)$:
\begin{equation}\label{eq:local_glob_rel}
     \frac{1}{\sqrt{\xi'(x)}}\varphi_{\pm , 0, \infty}\left(\xi(x)\right) = \exp[\pm  C(\hbar)]\phi_{\pm , 1}(x)
\end{equation}
where
\begin{equation}
\begin{aligned}\label{eq:definition-of-C}
     C(\hbar) &= \log\left( \frac{\varphi_{+, 0, \infty}(\xi(x))}{\phi_{+, 1}(x)\sqrt{\xi'(x)}} \right)\\
     &=\left(\frac{1}{\hbar}\int_0^{\xi(x)}\tilde S_{\text{odd},-1}(\xi')\, d\xi' + \int_\infty^{\xi(x)}\tilde S_{\text{odd, reg}}(\xi')\, d\xi' +\right)- \frac{1}{2}\int_{\gamma_{1,x}} S_{\text{odd}}(x')\, dx'\,.
\end{aligned}
\end{equation}
The quantity $\tilde S_\text{odd}(\xi)$ is as defined in \eqref{eq:sodd_def} but with the potential \eqref{eq:bessel_Potential}. 

Using \eqref{eq:local_WKB_sol_bes} and \eqref{eq:local_glob_rel} we deduce that the linear combination of WKB solutions satisfying the boundary condition at $x =0$ is 
\begin{equation}\label{eq:lin_com_zero_LC}
    \rme^{- \rmi p \pi /2 + 2C}\phi_{+, 1}(x) + \phi_{-,1}(x)\,.
\end{equation}

We can now derive the (perturbative) quantisation condition obeyed by $C(\hbar)$. To satisfy the boundary condition at $x= +\infty$, we start with $\phi_{-, 1}(x)$ near $x =+\infty$. We then analytically continue $\phi_{-, 1}(x)$ near to $x =0$. If we do this, the connection formula \eqref{eq:connection_form_LC} tells us that 
\begin{equation}
    \phi_{-,1}(x) \to   \phi_{-,1}(x) +\rmi \phi_{+,1}(x)\,.
\end{equation}
For the boundary condition at $x =0$ to be satisfied we require that close to zero, we have the linear combination \eqref{eq:lin_com_zero_LC}. This can only occur if 
\begin{equation}
    \rmi = \rme^{-\rmi p \pi/2 +2 C},
\end{equation}
or, by taking the logarithm,
\begin{equation}\label{eq:Chbar-quantised}
    C(\hbar) = \pi \rmi\left( n +
    \frac{p+1}{4}\right), \quad n \in \mathbb{Z}.
\end{equation}
This is the perturbative quantisation condition obeyed by $C(\hbar)$. The explicit calculation of $C(\hbar)$ as a series in  $\hbar$ is discussed in appendix \ref{ap:large_n_dets}. Following the discussion in the appendix, we compute the leading term small-$\hbar$ term of expansion of $C(\hbar)$ and find that the leading order quantisation condition is 
\begin{equation} \label{eq:leading_ord_qc}
    \frac{1}{\hbar} \frac{\rmi }{p+1}B\left(\frac{3}{2}, \frac{1}{1+p}\right) + \mathcal{O}(\hbar)= \pi \rmi \left(n + \frac{p+1}{4}\right)\,,
\end{equation}
where $B(x,y)$ is the Beta function. Since $\hbar>0$ is small, we must have $n+\frac{p+1}{4}>0$ and sufficiently large. Rearranging the leading order quantisation condition \eqref{eq:leading_ord_qc} we find that the large-$n$ formula for $k_{0}$ is\begin{equation} \label{eq:leading_order_kappa_n}
    k_{0}= \left(\frac{ 2\pi }{B\left(\frac{3}{2}, \frac{1}{1+p}\right)}\left( n +\frac{p+1}{4}\right) \right)^{\frac{2(p+1)}{p+3}}\left[1+ \mathcal{O}\left(\left(n+\frac{p+1}{4}\right)^{-2}\right)\right]
\end{equation}
for $p\geq 2$ and $n$ a non-negative integer. The large-$n$ estimates compared to the numerical values of \eqref{eq:k_0_num} for $n=0$ are shown in Table \ref{tab:k0_comparison} and we find that the leading order large-$n$\footnote{By large-$n$ we mean large $n+(p+1)/4$.} result provides good agreement even for the lowest overtone $n=0$. As $p$ increases, one might expect more accurate results because the expansion parameter is $n+(p+1)/4$. However, this is too naive, as the coefficients of the large-$n$ expansion depend on $p$. 
\begin{table}[]
    \centering
    \begin{tabular}{|c|c|c|c|}\hline
         $p$& Large-$n$ $k_0$, evaluated at $n=0$ & Numerical $k_0$ for $n=0$ & Relative difference\\\hline
         $2$& $2.1154$&$2.1212$& $0.27\%$\\\hline
          $3$&  $2.1851$&$2.1407$&$2.07\%$\\\hline
         $4$&  $2.2317$& $2.1196$&$5.29\%$\\\hline
          $5$&  $2.2651$& $2.0827$&$8.76\%$\\\hline
    \end{tabular}
    \caption{
    Comparison between the large-$n$ estimate of $k_0$ in eq.~\eqref{eq:leading_order_kappa_n}, evaluated at $n=0$, and the numerical value of $k_0$ in eq. \eqref{eq:k_0_num} for the lowest overtone $n=0$ (the shear mode). The leading-order large-$n$ approximation for $k_0$ works reasonably well even for $n=0$.
    }
    \label{tab:k0_comparison}
\end{table}
If we take $p=3$ and keep only the leading term in $n$ we get that \eqref{eq:sdef} becomes
\begin{equation}
    s(\eta) =1 - \frac{1}{2}\pi^{2/3} \left(\frac{ \Gamma(7/4)}{\Gamma(5/4)}(n+1)\right)^{4/3}\left(1+\mathcal{O}\left((n+1)^{-2}\right)\right)\eta^{-4/3}+ \mathcal{O}(\eta^{-8/3})
\end{equation}
When conventions are adjusted\footnote{If $q_F$, $\omega_F$, and $n_F$ denote the $q$, $\omega$ and $n$ of \cite{Fuini:2016qsc} then we have $\q = q_F/2 $, $\w = \omega_F/2$ and $n=n_F-1$.} this agrees with the estimate of the subleading correction for the shear modes at real $\q$ presented in \cite{Fuini:2016qsc}.

Before turning to an all-orders calculation of the large $\q=\rmi \eta$ expansion of $s(\eta)$ it is worth mentioning that one could obtain the leading order formula \eqref{eq:leading_order_kappa_n} by using a semi-classical Bohr-Sommerfeld quantisation together with a so-called Langer modification \cite{Berry:1972}. When $p$ is odd we would directly obtain the perturbative large-$n$ series for $k_{0}$ to all orders. However, for $p\geq4$ and even, the later terms in the large-$n$ expansion would be incorrect without using the uniform WKB method. Nevertheless, we now turn to a systematic approach to determine the large $\eta\equiv - \rmi\q$ 
expansion of the shear mode frequency avoiding the need to take the large overtone limit.

\section{All orders WKB for  \texorpdfstring{$p=3$}{p=3}}\label{sec:all_orders}

We now focus on how to retrieve large orders of the large-$\q$ expansion of the shear mode for imaginary $\q$. Because of its asymptotic nature, the large-$\q$ expansion will allow us to retrieve the small $\q$ behaviour of $\w(\q)$, and in particular obtain the shear viscosity, following the approach proposed in \cite{Aniceto:2026zmc}.

The calculation described in the last section can be generalised to find further terms of the large-$\q$ expansion of $\kappa(\q)$ as large-$n$ (overtone) series. This requires taking a double scaled limit as was done in \cite{Aniceto:2026zmc}. This derivation is limited by the number of terms we include in the large-$n$ expansion of each coefficient. Furthermore, we expect each coefficient to be a divergent series in $n$ that is supplemented by non-perturbative terms in $n$ that we ignored in the previous section. Such non-perturbative corrections can be found as shown in \cite{Aniceto:2026zmc} but obtaining accurate large-$\q$ coefficients at later orders requires an increasing number of terms in the large-$n$ expansions. 

These limitations motivate the use of a more direct method to retrieve the large-$\q$ coefficients of $\w$. 
As before we will be considering $\q$ imaginary,
$ \q =\rmi\eta$ with real $\eta$. 
In the following, we restrict ourselves to $p=3$, but the same method should work for general $p$ with suitable modifications. As discussed in section \ref{sec:rev-leading-all-p}, $\w(\eta)$ is given by
\begin{equation}\label{eq:omega-of-kappa}
    \w^2 =-\eta^2 \,\left(1-\rm\kappa(\eta)\eta^{-4/3}\right)
\end{equation}
with 
\begin{equation}\label{eq:kappa-of-omega}
    \kappa(\eta)=\sum_{n=0}^\infty k_{n}\eta^{-4n/3}.
\end{equation}
Note that we have not proved that the subleading powers of $\kappa(\eta)$ are $\eta^{-4/3}$. We treat this as an assumption that will prove to be correct. 

Since in this section we are considering $p=3$ which is odd, we can exploit an extra redundancy in the coordinate $u$ by setting $z = u^2$. Changing variables to use $z$ in \eqref{eq:Srchodinger-eq-u} will introduce a $\phi'(z)$ term  which we remove by setting
\begin{equation}
    \phi(z) =z^{-1/4}\psi(z).
\end{equation}
The equation now reads
\begin{equation}\label{eq:zeqn}
    \psi''(z)-V(z) \psi(z) =0
\end{equation}
where 
\begin{equation}\label{eq:vz}
     V(z) = \frac{\eta^2 z}{(1-z^2)^2}
     -\frac{\eta^{2/3}\kappa(\eta)}{z(1-z^2)^2}-\frac{(3-2z^2)}{(1-z^2)^2}.
\end{equation}
In terms of the new coordinate $z$, the boundary conditions become 
\begin{equation}
    \psi(z) \sim\begin{cases} 
        z, &\quad z \to 0,\\
        (1-z)^{\frac{1}{2}- \frac{\rmi \w}{2}}, &\quad z\to1.
    \end{cases}
\end{equation}
 Our aim is to find the large-$\eta$ expansion of $\kappa(\eta)$ as in equation\eqref{eq:kappa-of-omega}. 

 In the previous section, we argued that no Stokes phenomena occur if we analytically continue the solution $\phi_-(u)$ from $u=1$ to the vicinity of $u=0$. This meant that the boundary value problem reduced to an inner-outer matching of the WKB solution $\phi_-(u)$ to the Frobenius expansion near $u=0$. This inner-outer matching can be done using any coordinate, including $z$. The boundary value problem now requires that $\psi(z) \sim z$ for $z \to 0$ and $\psi(z) \propto \psi_-(z)$ for $\eta$ large and positive as well as for $z$ small and positive. In section \ref{sec:large_n}, we solved this problem at leading order in $\eta$ by calculating a large-$n$ series. In the next subsection, we systematically solve it to any order we wish using the uniform WKB method.

\subsection{Large-\texorpdfstring{$\q$}{q} QNMs from uniform WKB} 

Our aim now is to find a solution $\psi(z)$ that is decaying for large-$\eta$ close to $z = 0$ but regular at $z=0$. To find such a solution, we use the uniform WKB method, described in Section \ref{sec:rev-uniform-WKB}. 
 The first step of this method is to find a local potential $V_L(\xi)$ such that the generalised coordinate obeys $\xi(0)=0$. In section \ref{sec:rev-uniform-WKB}, we argued that the local potential $V_L(\xi)$ must at least match the leading small-$z$ power of each $\q$ or $\eta$ coefficient of $V(z)$.  Expanding each of the three terms in $V(z)$ in equation~\eqref{eq:vz} for small $z$, we find 
\begin{equation}
    V(z) = \eta^2{(z+\dots)} -\eta^{2/3}\left(\frac{\kappa(\eta)}{z}+\dots\right)-(3+\dots)
\end{equation}
This leads to the choice of local potential
\begin{equation}\label{eq:Vlocal_z}
    V_L(\xi) = \eta^{2} \xi -\eta^{2/3}\frac{A(\eta)}{\xi}+B(\eta)
\end{equation}
where
\begin{equation}
    \begin{aligned}
        A(\eta) = a_0 +...,\\
        B(\eta)= b_0 +... .
    \end{aligned}
\end{equation}
and by the ellipses we mean possible sub-leading powers of $\eta$ that can be deduced by the matching procedure.
 
We look for $\xi(z)$ in the form of  a large-$\eta$ expansion (see equation~\eqref{eq:xi-expansion-q}):
\begin{equation}
    \xi(z) = \sum_{n=0}^\infty\,
\eta^{-2n/3}\xi_n(z)
\end{equation}
Each coefficient $\xi_n(z)$ will then obey a first order differential equation that requires boundary conditions to fully specify the problem. We select the boundary condition $\xi_n(0)=0$ so that the behaviour of $\psi(z)$ near $z=0$ is determined by the behaviour of $G(\xi)$, (see equation \eqref{liouville}), near $\xi=0$. 
To calculate $\xi_n(z)$, we expand in powers of $z$ as (see equation~\eqref{eq:xi-expansion-u}) and demand that $\xi_n(0)=0$:
\begin{equation}
\xi_n(z)=\sum_{m=1}^\infty c_{n,m}z^m 
\end{equation}
We demonstrate this procedure for the first few coefficients $\xi_n(z)$. First, we expand the generalised coordinate equation \eqref{eq:gen_coord_eqn} for large $\eta$ and equate terms with corresponding powers of $\eta$. The highest power that appears is $\eta^2$ and demanding that terms of this order cancel, leads to the following equation for $\xi_0(z)$
\begin{equation}
    \xi'_0(z)^2 \xi_0(z) = \frac{z}{(1-z^2)^2}.
\end{equation}
This equation can be solved to find a closed form expression for $\xi_0(z)$. However, it is both sufficient for our purposes and generalisable to subsequent coefficients if we compute the expansion of $\xi_0(z)$ around $z=0$. We find
\begin{equation}
    \xi_0(z) = z+ \frac{2}{7}z^3 + \mathcal{O}(z^5)
\end{equation}
The next power is $\eta^{4/3}$. Equating terms at this power, we find an equation for $\xi_1(z)$ that can be solved trivially to find $\xi_1(z) \equiv0$. The next non-trivial equation occurs at order $\eta^{2/3}$, equating these powers yields an equation for $\xi_2(z)$:
\begin{equation} \label{eq:xi_2_eqn}
    -\frac{k_0}{z(1-z^2)^2}+\left(\frac{a_0}{\xi_{0}(z)}- \xi_2(z)\right)\xi_0'(z)^2 - 2 \xi_0(z) \xi_0'(z)\xi'_2(z) = 0.
\end{equation}
Note that to obtain the above equation we used the expansion for $\kappa(\eta)$ in equation\eqref{eq:kappa-of-omega}. Again, we find the small-$z$ series of $\xi_2(z)$ such that $\xi_2(0)=0$. Substituting the small-$z$ expansion of $\xi_0(z)$ into equation \eqref{eq:xi_2_eqn} and demanding that every power of $z$ vanishes we obtain the following leading order equation:
\begin{equation}
    \frac{1}{z}(k_0 - a_0)+ \dots  = 0.
\end{equation}
This means that the matching procedure forces $k_0 =a_0$. The constants $c_{2, m}$ can then be obtained by requiring that higher order terms in the above expansion vanish. 

At order $\mathcal{O}(1)$ in the large-$\eta$ expansion of \eqref{eq:gen_coord_eqn}, the requirement that $\xi(0)=0$ enforces $b_0 = -15/7$. At the next order, $\mathcal{O}(\eta^{-2/3})$, we find that
the matching procedure requires 
\begin{equation} \label{eq:matching1}
    \frac{1}{z}\left(-\frac{4}{21}k_0^2 -k_1\right) +\dots=0.
\end{equation}
At this stage in the calculation, we are simply transforming our differential equation \eqref{eq:zeqn} into a differential equation with potential $V_L(\xi)$. The boundary conditions have not been enforced yet meaning $\kappa$ and its coefficients are not yet fixed. However, for equation \eqref{eq:matching1} to be satisfied, we would require a constraint on $k_1$ in terms of $k_0$. This issue is resolved if the next term in the large-$\eta$ expansion of $A(\eta)$ is $a_1\eta^{-4/3}$. Then $a_1$ appears in equation \eqref{eq:matching1} and we can continue with the matching procedure provided:
\begin{equation}
    a_1 = \frac{4}{21}k_0^2 + k_1.
\end{equation}
Continuing in this way, we find that
\begin{equation}\label{eq:AB-expansions}
    \begin{aligned}
        A(\eta)= \sum_{n=0}^\infty a_n(k_0, \dots, k_n) \eta^{-4n/3},\\
         B(\eta)= \sum_{n=0}^\infty b_n(k_0, \dots, k_n) \eta^{-4n/3}\,
        \end{aligned}
\end{equation}
and the matching procedure at each order in $\eta$ fixes $a_n$ and $b_n$ in terms of the spectral coefficients $k_0, \dots, k_n$ (which are the expansion coefficients of $\kappa(\eta)$). Note that this process will only be correct provided we include enough terms in the small-$z$ expansions of the previous coefficients $\xi_{m}(z)$, $m =0, \dots, n-1$, so that the leading term in the small-$z$ expansion is calculated correctly.

We can now express the solutions to the original equation in $z$, equation \eqref{eq:zeqn}, as 
\begin{equation}\label{eq:bvp_solution_G}
    \psi(z) = \frac{1}{\sqrt{\xi'(z)}}G(\xi(z))
\end{equation}
where we can calculate $\xi(z)$ to arbitrary order in both $\eta$ and $z$, and $G$ obeys
\begin{equation}\label{eq:uniform-G-with-AB}
    G''(\xi) - \left[\eta^2 \xi -\eta^{2/3}\frac{A(\eta)}{\xi}+B(\eta)\right]G(\xi) = 0
\end{equation}
with $A(\eta)$ and $B(\eta)$ being large-$\eta$ series whose coefficients are specified in terms of the $\kappa(\eta)$ expansion coefficients $k_j$.

If we find a relationship between $A(\eta)$ and $B(\eta)$ such that $G(\xi)$ is regular at $\xi=0$ but decaying for large-$\eta$, then $\psi(z)$ will obey the boundary conditions. This restriction on $A(\eta)$ and $B(\eta)$ will then fix the coefficients $k_n$ of $\kappa(\eta)$. Usually when using the uniform WKB method, the relationship between the small and large argument expansions of the function $G$ is known (e.g. when $G$ is a special function such as the parabolic cylinder function) and enforcing the boundary conditions is straightforward. However, in our case $G(\xi)$ obeys an equation for which the solutions are not known, and finding the relationship between $A(\eta)$ and $B(\eta)$ is a necessary step, which we now turn to.

\subsection{Imposing the boundary conditions}

To solve the boundary value problem, we need to find a solution $G(\xi)$ to \eqref{eq:uniform-G-with-AB} that is regular at $\xi=0$ and decaying for large-$\eta$. The first condition ensures that the small $z$ expansion of \eqref{eq:bvp_solution_G} has the correct behaviour and the second condition ensures that the large-$\eta$ expansion of \eqref{eq:bvp_solution_G} for $z\in (0,1)$ is exactly $\psi_-(z)$.

As we have seen in the previous sections, we cannot naively expand the large-$\eta$ expansion of $G(\xi)$ for small-$\xi$ and expect to see the correct small-$\xi$ behaviour of this solution. This can be seen explicitly if one constructs the large-$\eta$ WKB expansion of \eqref{eq:uniform-G-with-AB}. Alternatively, by looking at $V_L(\xi)$, we can expect the large-$\eta$ expansion to break down when the leading term $\eta^2 \xi$ is of the same order as the sub-leading term $\eta^{2/3}\xi^{-1}$. This occurs when $\xi \eta^{2/3} \sim 1$ and motivates introducing a rescaled coordinate
\begin{equation}
    x = \eta^{2/3}\xi.
\end{equation}
Large $x$ corresponds to $\xi\gg\eta^{-2/3}$ (the large-$\eta$ regime) and small $x$ corresponds to $\xi \ll \eta^{-2/3}$ (the small-$\xi$ regime). In terms of $x$, the local equation \eqref{eq:uniform-G-with-AB} is
\begin{equation}\label{eq:local_eqn_inner}
    G''(x) -\left[x-\frac{A}{x} +g\right]G(x) =0\,,
\end{equation}
where
\begin{equation}
    g (\eta) = B \eta^{-4/3}.
\end{equation}
In the limit $\eta\to\infty$, $A(\eta)$ and $B(\eta)$ are of order one, meaning $g$ is small compared to $A$. Therefore, it is natural to look for $A$ as a perturbative series in $g$. Our task is then to find coefficients $\alpha_n$
\begin{equation}\label{eq:A_expansion}
    A(g) = \sum_{n=0}^\infty\alpha_n g^n,
\end{equation}
such that 
\begin{equation} \label{eq:inner_boundary_conditions}
    G(x) \sim \begin{cases}
        x, &\quad x\to 0,\\
       x^{-1/4}\exp\left(-\frac{2}{3}x^{3/2}\right), &\quad x\to \infty.
    \end{cases}
\end{equation}
Say we can find numbers $\alpha_n$ such that $G(x)$ solves the boundary value problem defined by \eqref{eq:local_eqn_inner} and \eqref{eq:inner_boundary_conditions}. Then we can find an expression for the large-$\eta$ expansion of $A(\eta)$ in terms of $\alpha_n$ and the large-$\eta$ coefficients of $B$, $b_n$, by reinstating $g=B\eta^{-4/3}$ and expanding $B(\eta)$ for large-$\eta$. This gives two independent large-$\eta$ expansions for $A(\eta)$ with coefficients dependent on $k_n$: one from transforming the original equation \eqref{eq:zeqn} to the local equation \eqref{eq:uniform-G-with-AB} with the requirement that $\xi(0)=0$; and one from solving the boundary value problem \eqref{eq:local_eqn_inner} and \eqref{eq:inner_boundary_conditions}. Requiring that these two expansions are equal fixes the values of $k_n$.

We now discuss how to solve the boundary value problem \eqref{eq:local_eqn_inner} and \eqref{eq:inner_boundary_conditions}. At leading order as $\eta\to\infty$ one can neglect the term proportional to $g$. Then one finds a generalised eigenvalue problem which determines $\alpha_0$
\begin{equation}\label{eq:local_eqn_inner_lead}
    G''(x) -\left[x-\frac{\alpha_0}{x}\right]G(x) =0.
\end{equation}
Since we need to solve this boundary value problem to high precision, it is natural to employ a spectral method. Given that the domain is the half-line, a natural basis is provided by the Laguerre polynomials (see the Appendix \ref{ap:numerical_method}). By solving the boundary value problem one finds a set of eigenvalues $\alpha_0$. As argued in Section~\ref{sec:lc-scaling}, the lowest one corresponds to the shear mode, as is confirmed by the resummation in Section~\ref{sec:viscosity}. We focus on this mode;   higher overtones will not be considered here, but their large-$\eta$ expansions can also be deduced from this method.

To find higher orders in $g$ we combine this spectral method with Rayleigh-Schr\"odinger perturbation theory, treating $A$ as the energy and $g$ as a perturbation. We emphasise that this method exploits the fact that the boundary conditions \eqref{eq:inner_boundary_conditions} are Dirichlet, a fact that derives from taking $\q$ positive imaginary. The details of this scheme are discussed in Appendix \ref{ap:numerical_method}.

Using this method, we can find $\alpha_0,\dots, \alpha_{101}$ to approximately 100 digits of precision in a few hours on a laptop. To find $k_n$, we replace $B(\eta)$ with its large-$\eta$ expansion in \eqref{eq:A_expansion}, we find 
\begin{equation}\label{eq:A_expansion_1}
    A(\eta) = \sum_{n=0}^\infty \Tilde{a}_n(\alpha_0, \dots, \alpha_n, b_0, \dots, b_n)\eta^{-4n/3}.
\end{equation}
The first three terms of this expansion are: 
\begin{equation}
      \tilde{a}_0 = \alpha_0\,,\quad
      \tilde{a}_1 = b_0\alpha_1\,,\quad
      \tilde{a}_2 = \alpha_2b_0^2+ \alpha_1b_1 \,.
\end{equation}
Then requiring that the coefficients $a_n$ of the original large-$\eta$ expansion of $A$, \eqref{eq:AB-expansions}, obtained by the uniform WKB matching procedure are equal to $\tilde{a}_n$ for all $n$ yields the values for $k_n$. For example, enforcing this requirement for $n=0$ and $n=1$ gives:
\begin{equation}
\begin{aligned}
       k_0 &= \alpha_0, \\
    k_1 &= -\frac{15}{7}\alpha_1 -\frac{4}{21}k_0^2. 
\end{aligned}
\end{equation}
where the function $b_0(k_0)$ is known from the uniform WKB matching procedure. Continuing in this way, we can find $k_0, \dots, k_{100}$ with great accuracy limited only by the determination of $\alpha_n$. The first few coefficients are as follows:
\begin{equation}
\begin{aligned}
    & k_0 \approx  2.14070046874 , \\
    & k_1 \approx -2.39351019012 ,\\
    & k_2 \approx  0.917498463558 ,\\
    & k_3 \approx  1.24009530999 ,\\ 
    & k_4 \approx  -1.38744948786 ,\\ 
    & k_5 \approx -1.24331105990 ,
\end{aligned}
\end{equation}
though in practice we have evaluated $k_n$ to a significantly higher precision. We have thus determined the expansion of the shear-mode frequency at large $\q=\mathrm{i}\eta$, and will now discuss the consequences of this expansion.

\section{Viscosity from the large-\texorpdfstring{$\q$}{q} expansion}
\label{sec:viscosity}

 The large-$\q$ expansion of the frequency $\w(\q)$ \eqref{eq:omega-of-kappa}, determined with the tools presented above, is factorially divergent. It is well-established that sufficiently well-behaved divergent expansions, generally those that are of Gevrey-1 type \cite{Aniceto_2019}, can provide sensible results, and analytic information well beyond the na\"ively expected regime of applicability.  Then one naturally asks: can we analytically continue this large $\q$ expansion all the way into the hydrodynamic regime?
As shown in \cite{Aniceto:2026zmc}, this can indeed be done via a combination of Borel resummation up to a finite value of $\q$, followed by analytic continuation through Pad\'e approximants all the way to $\q=0$, which we will briefly summarise here. 

Using the methods discussed in the previous sections, we have determined 101 coefficients $k_n$ of the expansion in large $\q$ \eqref{eq:kappa-of-omega} appearing in the frequency $\w(\q)$ \eqref{eq:omega-of-kappa}.\footnote{The coefficients were determined with a precision of up to 100 digits.} This series is factorially divergent, up to non-perturbative (exponentially small) terms, and it represents the function $\w(\q)$ at finite values of $\q$ via resummation. Resurgence theory and Borel analysis \cite{Aniceto_2019} provide an approach to determine the existence of such non-perturbative corrections and the resummation procedure. A large order analysis of the coefficients of the expansion \eqref{eq:kappa-of-omega} shows that their growth is $k_n\sim \Gamma(4n/3)$. The resummation procedure then proceeds as follows. The first step is to determine the Borel transform $\mathcal{B}[\kappa](t)$ of the series \eqref{eq:kappa-of-omega} (recalling that $\eta=-\mathrm{i}\q$), which can be obtained by the map\footnote{This is equivalent to an inverse Laplace transform of each element of the expansion.} 
\begin{equation}
    \eta^{-\alpha}=(-\mathrm{i}\q)^{-\alpha}\rightarrow \frac{t^{\alpha-1}}{\Gamma(\alpha)}.
\end{equation} 
valid for $\alpha> 0$, where $t$ is the Borel-plane variable. Note that the first term of the series $\kappa(\eta)$ needs to be removed in the calculation of the Borel transform, and added during the process of resummation. 

\begin{figure}
    \centering
\includegraphics[width=0.6\linewidth]{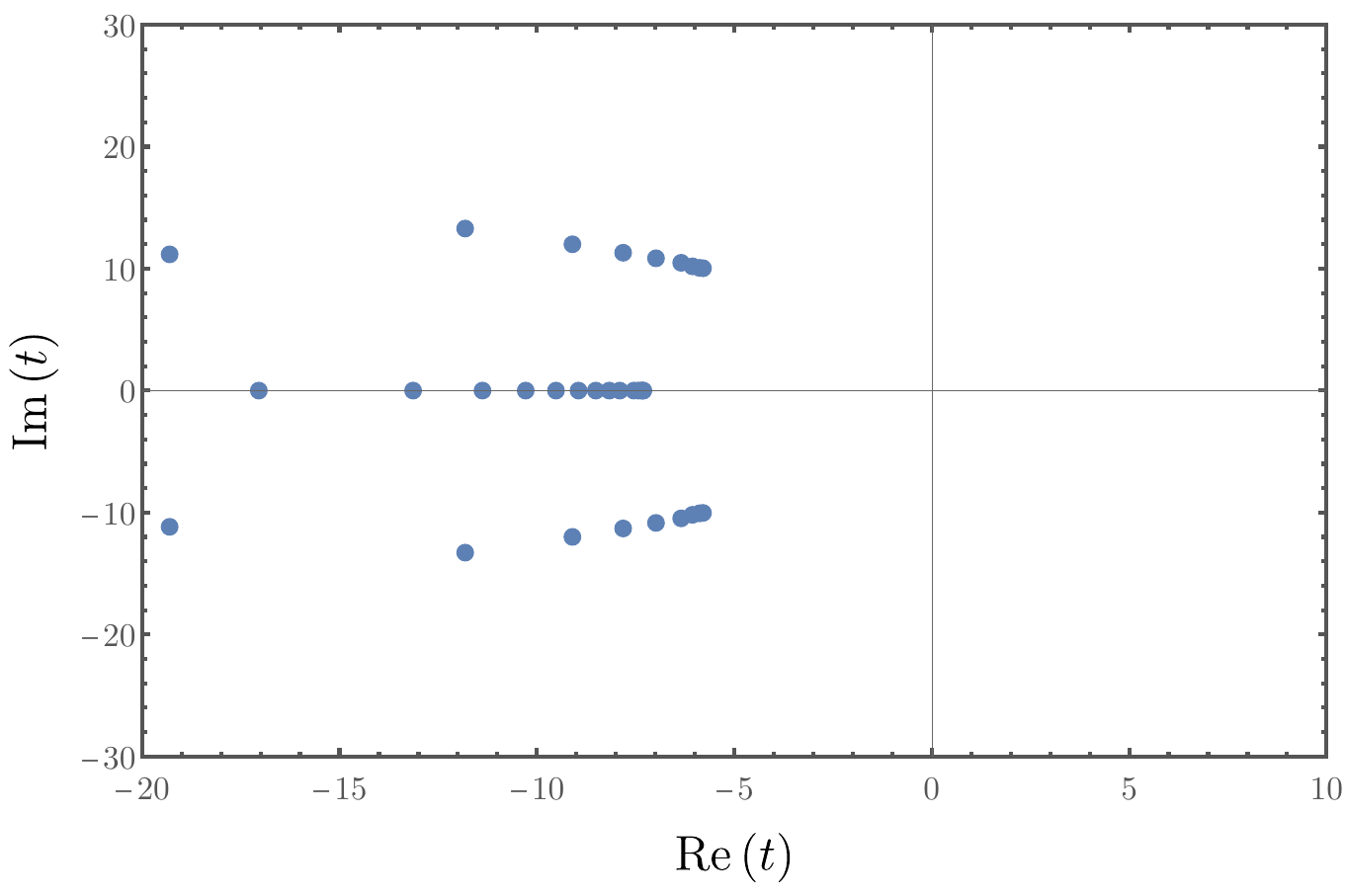}
    \caption{Poles of the Pad\'e approximant (diagonal, order 50) of the Borel transform $\mathrm{BP}_{50}[\kappa](t)$, associated to the large $\q=\mathrm{i}\eta$ expansion of $\kappa(\eta)$ in \eqref{eq:kappa-of-omega}.}
    \label{fig:Borel-sing}
\end{figure}

The Borel transform is now given by a series around the origin $t=0$ with a finite radius of convergence. We analytically continue
the Borel transform series 
through a (diagonal) Pad\'e approximant of order $L$ at $t=0$. Given a convergent power series $f(t)$ around a point $t_0$, a Pad\'e approximant $\mathrm{P}_L[f](t;t_0)$ of order $L$ at the same point takes $2L$ terms of the series and approximates it with a ratio of two polynomials of order $L$. The order-$L$ Pad\'e approximant of the Borel transform\footnote{The Borel transform is an expansion in integer powers of $t^{4/3}$ (up to an overall power of $t^{1/3}$). Its Pad\'{e} approximant is determined as a ratio of polynomials in this variable, not in $t$.} of $(\kappa(\eta)-k_0)$ will be denoted by
\begin{equation}
    \mathrm{BP}_L[\kappa](t) = {\rm P}_L[\mathcal{B}[\kappa]](t;0) \,.
\end{equation}
The position of the poles of the Borel-Pad\'e approximant $\mathrm{BP}_{50}[\kappa](t)$ is presented in Figure~\ref{fig:Borel-sing}. The accumulation of poles indicates the emergence of 3 branch cuts located in the left quadrants ($\mathrm{Re}(t)<0$). These cuts in the Borel plane of complex $t$ correspond to corrections that are non-perturbative in the large-$\eta$ expansion. Such corrections are expected to be of the form $\exp(A \eta)$ for some constant $A$ with positive real part. The derivation of the series $\kappa(\eta)$ discussed in Sections \ref{sec:lc-1} and \ref{sec:all_orders} determined that there are no such non-perturbative contributions to $\kappa(\eta)$ for real positive $\eta$.\footnote{The Borel plane picture is consistent with this: at $\mathrm{Re}(\eta)>0$ the non-perturbative contributions, if they were present, would be exponentially enhanced.} Given this fact, the branch cuts in the Borel plane show that non-perturbative terms in $\eta$ would be present if we analytically continued $\kappa(\eta)$ to $|{\rm arg}(\eta)|>\theta_c$, where $\theta_c$ is the argument of the location of the branch point in the upper left quadrant of Figure \ref{fig:Borel-sing}.

The resummed function $\kappa (\eta_0)$ at the finite real value of $\eta_0$ is then given by the Laplace transform 
\begin{equation}\label{eq:kappa-resummed}
\mathcal{S}\kappa_L(\eta_0)=k_0+\int_0^{+\infty}\mathrm{d}t \,  \mathrm{e}^{-t\,\eta_0}\,\mathrm{BP}_L[\kappa](t)\,
\end{equation}
which corresponds to the resummed approximation obtained from the $2L$ coefficients of the large-$\eta$ expansion \eqref{eq:kappa-of-omega}.
The value of the $m^{\mathrm{th}}$ derivative of $\kappa(\eta)$ at $\eta=\eta_0$ is equally simple and is given by
\begin{equation}\label{eq:kappa-deriv-resummed}
\mathcal{S}\kappa_L^{(m)}(\eta_0)=\int_0^{+\infty}\mathrm{d}t\,  \mathrm{e}^{-t\,\eta_0}\,(-t)^m\,\mathrm{BP}_L[\kappa](t), \quad m\geq 1\,.
\end{equation}
Starting with total number of terms $2L=100$ we have determined up to $m_{\mathrm{max}}=19$ derivatives at a range of finite reference values of $\eta_0\in\{\frac{3}{2}, 2, \frac{5}{2}, 3,\frac{7}{2},4\}$ -- chosen to be small enough to be close to $\q=0$, but also large enough that the error in the Borel resummation is not large. In order to ascertain the accuracy of the derivatives we varied the number of initial terms $2L$ between $L=48$ and $L=50$, and verified the stability of the final result: the relative difference of leading value $m=0$ varied between $10^{-12}$ for $\eta_0=\frac{3}{2}$ and $10^{-29}$ for $\eta_0=4$; the relative error of, e.g., the $m=7$-th derivative varied between $10^{-4}$ for $\eta_0=\frac{3}{2}$ and $10^{-21}$; the relative error of the last derivative calculated $m=19$ varied between $10^{-2}$ for $\eta_0=\frac{3}{2}$ and $10^{-13}$. 
We can thus obtain the analytic behaviour of  $\kappa(\q=\mathrm{i}\eta)$ around each of the values $\eta_0\in\{\frac{3}{2},2,\frac{5}{2},3,\frac{7}{2},4\}$ as a Taylor series with $m_{\max}=19$ terms
\begin{equation}\label{eq:Taylor-kappa-at-eta0}
    \kappa_{L}(\eta;\eta_0)=\sum_{m=0}^{m_{\max}}\,\frac{\mathcal{S}\kappa_L^{(m)}(\eta_0)}{m!}\,(\eta-\eta_0)^m\,.
\end{equation}
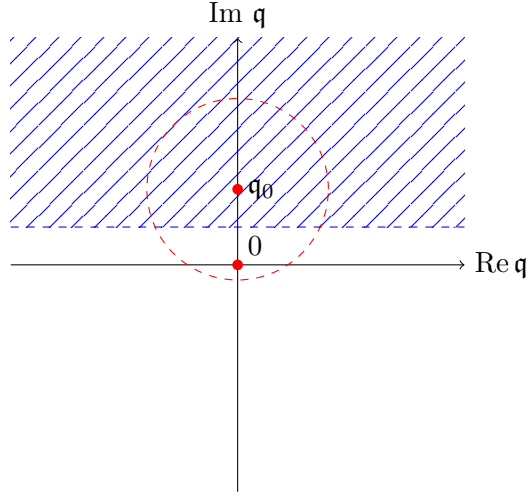
\begin{figure}
    \centering
    \begin{tikzpicture}
 
  \draw[->] (-5,0) -- (1,0) node[right] {$\mathrm{Re}\, \mathfrak{q}$};
\draw[dashed, blue] (-5,0.5) -- (1,0.5)  ;
 \fill[
    pattern={Lines[angle=45, distance=6pt]},
    pattern color=blue
  ] (-5,0.5) rectangle (1,3);

  \draw[->] (-2,-3) -- (-2,3) node[above] {Im $\mathfrak{q}$};
  \fill[red] (-2,1) circle (2pt) node[right, black] {$\mathfrak{q}_0$};
  \draw[dashed,red] (-2,1) circle (1.2); 
   \fill[red] (-2,0) circle (2pt) node[above right, black] {$0$};
\end{tikzpicture}
    \caption{
    A schematic representation of the process used to calculate the small-$\q$ expansion of $\w(\q)$ from its large-$\q$ expansion. The blue shaded area represents a region of the complex $\q$ plane where the resummation via the inverse Laplace transform~\eqref{eq:kappa-resummed} is valid. The dashed red circle represents the validity of the Taylor series of the function $\kappa(\eta)$ around $\eta_0 = -i \q_0$, whose coefficients were obtained from the large-$\q$ expansion using eq.~\eqref{eq:kappa-deriv-resummed}. 
    }
    \label{fig:placeholder}
\end{figure}
We are interested in the behaviour of the shear-mode frequency near $\q=0$. However, the Laplace integrals of Borel resummation are not defined at that point and the integral converges much slower at values of $\q$ that are very small (for the same number of terms at large $\q$). Instead, we use the analytic information in the form of the Taylor series of $\kappa_L(\eta;\eta_0)$ at each  point $\eta_0$ (which has a finite radius of convergence) to determine a Pad\'e approximant at that point and then use it to analytically continue to $\q=0$. We can choose the order of the (diagonal) approximant up to $9$ as we have 20 terms in the Taylor series, but one can choose a smaller order to make sure that the derivatives that are included have a smaller associated error.\footnote{Note that for the Pad\'e approximant we assumed  $\w$ as a power series in $\q^2$, motivated by the form of the original ODE. If one had 
instead assumed a series in powers of $\q$, the coefficients would not be stable when changing the reference points $\eta_0$.} We consider the whole range of reference values $\eta_0\in\{\frac{3}{2},2,\frac{5}{2},3,\frac{7}{2},4\}$ and for each we determine the diagonal Pad\'e approximants of orders from $4$ to $9$: we determine $\mathrm{P}_N[\w_{50}](\eta;\eta_0)$ for $N=4,\cdots,9$ and $\eta_0\in\{\frac{3}{2},2,\frac{5}{2},3,\frac{7}{2},4\}$. We then expand the resulting frequency \eqref{eq:omega-of-kappa} as a series in $\eta$ about $\eta=0$. Since we do not know \textit{a priori} which reference value $\eta_0$ and order of Pad\'e approximant $L$ provide the best approximation, we perform an average over all the expansions at $\q=0$, and obtain (for $2L=100$ original terms, $L=50$):
\begin{equation}
    \w_{50}(\q)\sim -4.96397\times 10^{-6}\rmi-0.49998\mathrm{i}\,\q^2-0.07617\mathrm{i}\,\q^4-0.01500\mathrm{i}\,\q^6+\mathcal{O}(\q^8)\,.
\end{equation}
The shear mode's behaviour at small $\q$ has been determined analytically to be \cite{Grozdanov:2019uhi}
\begin{equation}
    \w(\q)=-\frac{\mathrm{i}}{2}\,\q^2-\frac{1}{4} (1-\log (2))\mathrm{i}\,\q^4-\frac{1}{96} \left(24 \log ^2(2)-\pi ^2\right)\mathrm{i}\,\q^6+\cdots\,.
\end{equation}
Comparing the two results above shows exceptionally good agreement of the coefficient of the leading term (corresponding to the shear viscosity to entropy density ratio $\eta/s=1/4\pi$)  with relative error of $3.4\times10^{-5}$, and of the next term with error of $0.71\%$.

One could ask whether we could have a similar success starting with just a small number of terms of the large $\q$ expansion, such as $2L=10$ . In this case an analysis of the relative error of the derivatives shows that only the reference values $\eta_0\in\{\frac{7}{2},4\}$ and the first 10 derivatives are reliable to use in the analytic continuation. Performing the same computation (with the last Pad\'e approximants being only of order $4,5$) we then obtain the values:
\begin{equation}
    \w_{5}(\q)\sim 0.00126\rmi-0.49774\mathrm{i}\,\q^2+\cdots\,.
\end{equation}
Even when using just 10 terms from the large $\q$ expansion, one still obtains a prediction of the viscosity with an error of $0.45\%$.

\begin{figure}
\centering
\includegraphics[width=0.6\textwidth]{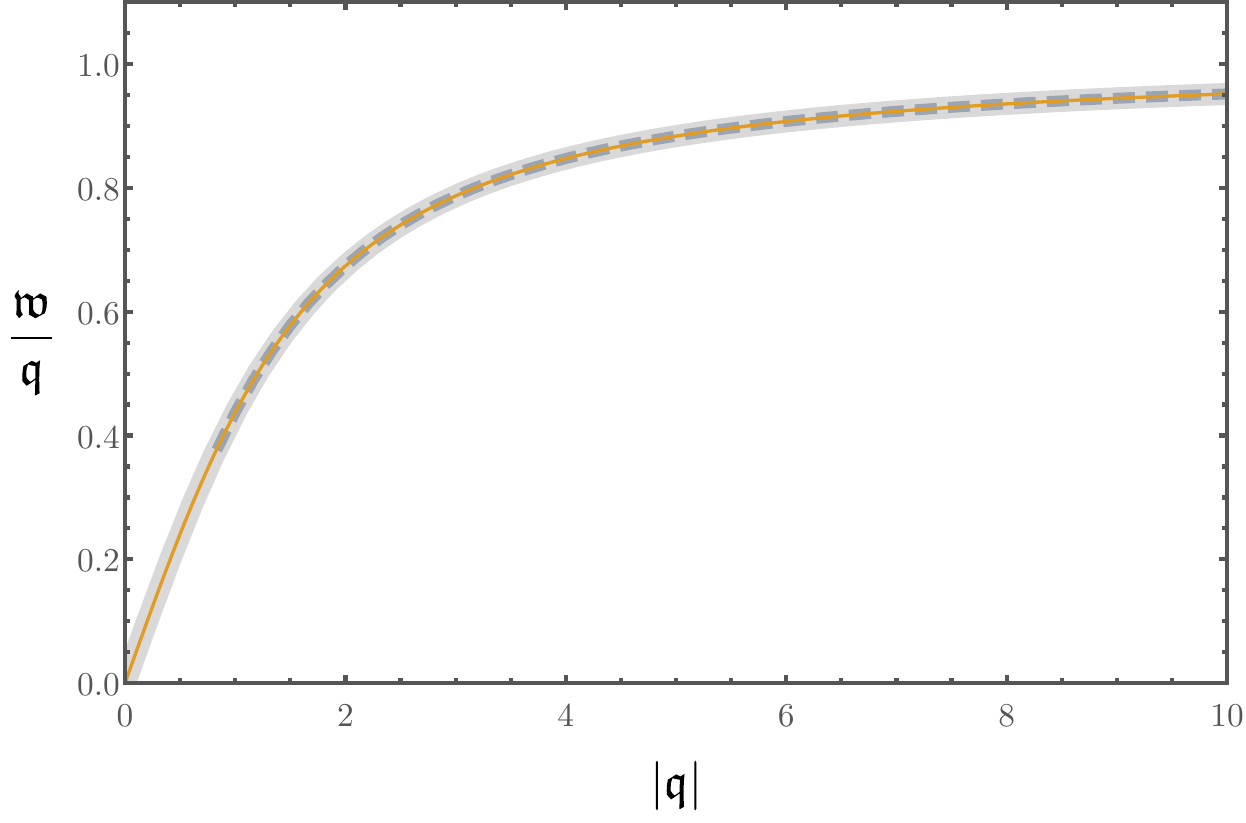}
\caption{
\label{fig:comparison-shear-mode}
Shear mode frequency $\w(\q)$ for AdS$_{5}$ black branes~\eqref{eq:bb0}, at positive imaginary~$\q$. The orange line is the same orange line as in Figure~\ref{fig:shear-mode-imq}, corresponding to the numerical calculation of $\w(\q)$. The darker dashed line corresponds to the frequency \eqref{eq:omega-of-kappa}  with $\kappa(|\q|)$ determined as a Borel resummation of the same $100$-term large-$|\q|$ expansion (as in \eqref{eq:kappa-resummed}) at the values of $|\q|\geq 5/6$, and $L=50$. The thick gray line corresponds to a diagonal Padé approximant of order $9$ associated to the Taylor expansion \eqref{eq:Taylor-kappa-at-eta0} at $|\q|=5/2$. The lines are drawn thick for visualisation only; their thickness does not indicate a numerical error or uncertainty. 
}
\end{figure}

Plotted in Figure \ref{fig:comparison-shear-mode} is the shear mode frequency obtained in 3 different ways. The results obtained from resummations of the large $\q$ expansion of $\kappa(\eta)$ (including $101$ terms) are given by the gray band and the thick dashed line, while the numerical evaluation of Figure \ref{fig:shear-mode-imq} for $p=3$ is given by the thin orange line. The thick dashed line is the result of evaluating the frequency \eqref{eq:omega-of-kappa} via direct Borel resummation \eqref{eq:kappa-resummed} for various values of $\eta\ge5/6$, and $L=50$.\footnote{For smaller values of $\eta$ the convergence of the Laplace integral is too slow, given the number of terms $2L=100$ of the large $\q$ expansion.} 
The gray band is a plot of the Padé approximant $\mathrm{P}_9[\w_{50}](\eta)$ of order $9$ determined from the Taylor series \eqref{eq:Taylor-kappa-at-eta0}, determined at the reference value $\eta_0=5/2$.  We can see that there is clear agreement between all three.

\section{Discussion}
\label{sec:Discussion}

At finite temperature, singularities of real-time response functions of local operators contain physical information about excitations of the macroscopic system and its approach to equilibrium. A convenient way to characterize these  singularities is through their dispersion relations $\w(\q)$ in the Fourier space. For the response functions of the energy-momentum tensor, the two classes of the most physically identifiable singularities are a) gapless long-distance singularities, for which the momentum $\q\to0$, with the associated frequency $\w(\q)\to0$, and b) light-cone short-distance singularities, for which the momentum $|\q|\to\infty$, with the associated frequency $\w(\q)\sim\pm\q$. The former connect to the familiar physics of hydrodynamics and macroscopic transport, while the latter connect to the ultraviolet structure of the underlying relativistic field theory. Both the physical interpretation and the mathematical methods used to study the two regimes are typically very different. In field theories which admit a holographic description in terms of classical gravity (such as ${\cal N}=4$ supersymmetric Yang-Mills theory at large $N_c$ and large 't~Hooft coupling), the above two limits of $\w(\q)$ can be continuously connected by changing $\q$. As a result, macroscopic hydrodynamic excitations can be directly traced to their fundamental microscopic origin. In this paper, we focus on the shear mode (excitations of transverse momentum density) in holographic theories where $\w(\q)$ can be computed from Einstein gravity in AdS$_{p+2}$. 

Our main technical result is the  representation of the shear mode frequency as an asymptotic series 
valid at large $|\q|$.
We begin by summarising those results and their implications for the case $p=3$, which corresponds to ${\cal N}=4$ supersymmetric Yang-Mills theory in 4-dimensional Minkowski space. 
The shear-mode frequency can be expressed as an asymptotic series in powers of $1/|\q|^{4/3}$, which we repeat here for clarity. For positive imaginary $\q = \rmi \eta$, the expansion is
\begin{equation}
    \w^2 =-\eta^2 \,\left(1-\rm\kappa(\eta)\eta^{-4/3}\right)
\end{equation}
with 
\begin{equation}
\label{eq:kappa-of-q}
    \kappa(\eta)=\sum_{n=0}^\infty k_{n}\eta^{-4n/3}.
\end{equation}
The method of evaluating the coefficients $k_n$ is described in Section \ref{sec:all_orders}. It entails using an all-orders WKB method where $1/|\q|$ plays the role of the Planck constant. 
To get the coefficients of this series beyond the leading order we needed to employ WKB methods beyond the leading semi-classical approximation, including a uniform WKB map connecting the boundary of $AdS$ to the black brane horizon. 
The series \eqref{eq:kappa-of-q} has a vanishing radius of convergence, but can be resummed using the  Borel technique, as explained in Section \ref{sec:viscosity}. Furthermore, informed by a resurgence analysis, we showed that $\w(\q)$ does not have any non-perturbative corrections for $\q$ positive imaginary. The resummed function yields detailed information about $\w(\q)$ in a neighbourhood of the imaginary $\q$-axis all the way to the vicinity of the origin. In practice, only a finite number of terms of the large-$\q$ series are available -- we have obtained 101. A workable approach is to evaluate $\w(\q)$ along with some number of its derivatives at a fixed point $\q=\q_0$ with $|\q_0|$ of order one. The knowledge of the function's derivatives at $\q_0$ opens the way to analytically continuing the quasinormal mode frequency all the way from $\q_0$ to the origin. One way to proceed is to  calculate the Pad\'e approximant to the Taylor series at $\q_0$. The Pad\'e approximant turns out to have no singularities at (or near) $\q=0$, which strongly suggests that $\w(\q)$ is analytic there. 
Expanding the Pad\'e approximant at small $\q$ shows that $\w(\q)$ vanishes at $\q=0$ to high precision. These properties are tantamount to the statement that hydrodynamics emerges in the long distance limit. 

Calculating the leading terms of the Taylor series around $\q=0$ by using 101 terms in the large-$\q$ expansion, we find that the coefficient of the $\q^2$ term matches the expected value of the shear viscosity to entropy density ratio $\eta/s=1/4\pi$ to one part in $25 000$. Higher-order terms in the hydrodynamic small-$\q$ expansion of $\w(\q)$ can also be obtained, though with less precision.

In our view, access to hydrodynamic transport coefficients from the ``ultraviolet'' large-$\q$ data is a fascinating connection which works better than one might naively have guessed. From a computational perspective, evaluating hydrodynamic transport coefficients in holography is of course more straightforwardly performed using the small-$\q$ expansion directly~\cite{Policastro:2002se}. However, from the perspective of quantum field theory, the small-$\q$ holographic calculation of transport coefficients is a ``black box'', with little field-theoretic intuition, and no visible connections to field-theoretic (ultraviolet) data. The results of this paper present a route towards establishing such connections.

Straightforward extensions of our work include the application of our methods to the diffusion of R-charge, connecting large-$\q$ and small-$\q$ regimes in other dimensions, and in theories beyond Einstein gravity (the near light-cone expansion at imaginary $\q$ was recently considered in theories beyond Einstein gravity in \cite{Buchel:2026rep, Buchel:2026ajv}).

We have also established a number of results valid for more general values of $p$. In fact, Section \ref{sec:lc-1} presents an all $p\ge2$ result for the sub-leading correction to the complex light-cone scaling. It is worth noting that the method used in Section \ref{sec:lc-1}  to determine this first correction to the complex light-cone scaling relied on (sub-)leading order WKB approximations, which did not require the sophistication introduced in Section~\ref{sec:all_orders}. However, such leading order methods are not sufficient to determine higher orders in the large-$\q$ expansion of $\w(\q)$. 
The analysis of the large-$\q$ expansion of the shear mode presented in Section \ref{sec:all_orders} can be extended to other overtones and other values of $p$.

Finally, it is intriguing to explore how the results of our paper may extend beyond simple holographic setups. Indeed, consider the retarded two-point function of helicity-1 components of the energy-momentum tensor in equilibrium, say $G^{tx,tx}(\omega, k)$ in 3+1 dimensional field theory, with $k$ in the $z$-direction. At real $k$, the singularities of the retarded function $G(\omega, k)$ lie in the lower half-plane of complex $\omega$. Let us assume that in that theory, hydrodynamics provides a valid low-energy effective description, in other words assume that at sufficiently small real $k$, there exists a separation of scales, such that the singularities of $G(\omega,k)$ near $\omega=0$ (hydrodynamic regime) are separated from the rest of the singularities further down in the lower complex $\omega$-plane by a finite gap (non-hydrodynamic excitations). What happens as $k^2$ changes from small positive values to small negative values? The non-hydrodynamic singularities will shift a little around their gap scale, while the shear diffusion mode $\omega \sim -i {\cal D} k^2$ will move into the upper half-plane, becoming the singularity with the largest ${\rm Im}(\omega)$. As $k^2$ decreases further, it is conceivable that the shear mode will stay the top-most mode on the imaginary axis, and will travel to $\omega\to i \infty$ as $k^2 \to-\infty$ unobstructed, just like it does in holographic theories studied here. If the field theory has an ultraviolet fixed point, a near light-cone expansion of $\omega(k)$ at large imaginary $k$ may give access to the hydrodynamic regime, similar to what happens in holography. We plan to explore such ``UV-IR'' connections in the future. 

\acknowledgments 

IA was partially supported by a UKRI EPSRC Early Career Fellowship EP/S004076/1. AR is supported by UKRI STFC consolidated grant ST/X000583/1, “New Frontiers in Particle Physics, Cosmology and Gravity”. MS is supported by the National Science Centre, Poland, under Grants No. 2021/41/B/ST2/02909 and 2025/59/B/ST2/02995. PK would like to thank Andrei Parnachev for helpful conversations.

\appendix

\section{The large overtone limit}\label{ap:large_n_dets}

In this appendix we review some of the technical details that are needed to obtain the large overtone expressions presented in Section \ref{sec:large_n}. 

To determine the linear combination of WKB solutions that satisfy the boundary condition at $x = 0$ one needs to determine $C(\hbar)$ (defined in equation \eqref{eq:definition-of-C}) as a small-$\hbar$ series, which is then quantised as in \eqref{eq:Chbar-quantised}.

In order to calculate the series expansion for $C(\hbar)$, we first calculate the integral 
\begin{equation}\label{eq:Sodd_int}
    \int_{\gamma_{1,x}}S_\text{odd}(x') \, dx'.
\end{equation}
 The small $\hbar$ expansion of $S_\text{odd}$ derives from the recursion relation \eqref{eq:riccati_perturb} with $V_0(x)$ and $V_2(x)$ taken from \eqref{eqn:LC_inner_leading_eqn}. The first two terms are
\begin{equation}
\begin{aligned}
      S_{\text{odd},-1}(x) &= \sqrt{x^{1+p}-1},\\
      S_{\text{odd, 1}}(x) &= x^{-2}\left(x^{1+p}-1\right)^{-5/2}\left(\frac{p(p-2)}{8} -\frac{3 p(p-1)}{8}x^{1+p}+\frac{(p-5)(1+3p)}{32}x^{2+2p} \right).
\end{aligned} 
\end{equation}
By induction and using the recursive definition \eqref{eq:riccati_perturb}, one can show that for $n \ge 0$ $S_{\text{odd},2n-1}(x)$ is of the form 
\begin{equation}
    S_{\text{odd}, 2n-1 } = x^{-2n} \left(x^{1+p}-1\right)^{-3n+\frac{1}{2}}\sum_{k=0}^{2n} R_{n,k}(p) x^{k(1+p)}
\end{equation}
for some polynomials in $p$, $R_{n, k}(p)$. This means that the integral of $S_\text{odd,2n-1}(x)$ around the branch cut is just a linear combination of integrals of the form
\begin{equation}
    \int_{\gamma_{1,x}} \left(z^{1+p}-1\right)^{-3n+1/2}z^{-2n+ k(1+p)}\, dz = \frac{2 \rmi(-1)^{n+1} }{p+1} B_{1-x^{1+p}}\left(\frac{3}{2} -3n, k +\frac{1-2n}{1+p}\right).
\end{equation}
where $B_x(a, b)$ is the incomplete Beta function. Therefore, the coefficients of the small-$\hbar$ expansion of $\eqref{eq:Sodd_int}$ are given by 
\begin{equation}\label{eq:Sodd_int_coeff}
    \int_{\gamma_{1,x}} S_{\text{odd}, 2n-1}(x') \, dx' = \frac{2\rmi (-1)^{n+1}}{p+1}\sum_{k=0}^{2n}R_{n, k}(p) B_{1-x^{1+p}}\left(\frac{3}{2} -3n , k + \frac{1-2n}{1+p}\right).
\end{equation}
The quantity $C(\hbar)$ is a $\hbar$ dependent proportionality constant between two solutions of the differential equation \eqref{eqn:LC_inner_leading_eqn} so must be independent of $x$. This means that any $x$ dependence in \eqref{eq:Sodd_int_coeff} must be cancelled by the other terms included in $C(\hbar)$: the integrals of $\tilde S_{\text{odd}}(\xi(x))$. We note that these integrals can themselves contain constant terms for $n\geq1$ and one would need to include such terms to go beyond leading order.

To extract the constant part of \eqref{eq:Sodd_int_coeff}, we expand for $x$ close to 0. Provided $k + \frac{1-2n}{1+p} \neq -1, -2, \dots$\footnote{For even $p$ and $n\geq1$ we can have $k+\frac{1-2n}{1+p}=-1,-2,\dots$. In this case, the constant term can be extracted by considering $k+\frac{1-2n}{1+p}+\epsilon$ and taking the limit $\epsilon\to 0$.} we have 
\begin{equation}\label{eq:incomplete_beta_func}
     B_{1-x^{1+p}}\left(\frac{3}{2} -3n , k + \frac{1-2n}{1+p}\right) = B\left(\frac{3}{2} -3n , k + \frac{1-2n}{1+p}\right)-  B_{x^{1+p}}\left(\frac{3}{2} -3n , k + \frac{1-2n}{1+p}\right)
\end{equation}
and the constant term is given by a linear combination of Beta functions. Thus, at leading order, the quantisation condition \eqref{eq:Chbar-quantised} is
\begin{equation}
    \frac{1}{\hbar} \left[\frac{\rmi }{p+1}B\left(\frac{3}{2}, \frac{1}{1+p}\right)\right] = \pi \rmi \left(n + \frac{p+1}{4}\right)\,.
\end{equation}

\section{Solving the generalised eigenvalue problem using spectral methods}\label{ap:numerical_method}

\subsection{General Method} \label{ap:general_meth}
In this section, we discuss the spectral method used to solve the boundary value problem \eqref{eq:local_eqn_inner}, \eqref{eq:inner_boundary_conditions} and how to find the coefficients of the expansion \eqref{eq:A_expansion}:
\begin{equation}
    A(g) =\sum_{n=0}^\infty\alpha_n g^n.
\end{equation}
First, we select $M$ basis functions $\{\phi_0(x), \dots, \phi_{M-1}(x)\}$ that vanish at $x =0$ and $x = \infty$. We then introduce the matrices $\textbf{H}$, $\textbf{V}$, $\textbf{W}$ whose components are defined as follows:
\begin{equation}\label{eq:appB-matrices-def}
    \begin{aligned}
        H_{i,j} &= \int^\infty_0\,\left(\phi_i'(x)\,\phi'_j(x) + x\phi_i(x) \,\phi_j(x)\,\right)\,dx,\\
         V_{i,j} &= \int^\infty_0\phi_i(x) \,\phi_j(x)\,dx,\\
          W_{i,j} &= \int^\infty_0\frac{ \phi_i(x) \,\phi_j(x)}{x}\,dx.
    \end{aligned}
\end{equation}
In terms of these matrices, the original eigenvalue problem \eqref{eq:local_eqn_inner} can be written as 
\begin{equation}\label{eq:matrix_inner}
    (\textbf{H} +g\,\textbf{V})\,\textbf{c}(g) = A(g) \,\textbf{W}\, \textbf{c}(g)\,
\end{equation}
where 
\begin{equation}\label{eq:appB-vector-def}
 \textbf{c}(g) =(c_0(g), \dots, c_{M-1}(g))^T\,.
 \end{equation}
 The solution to the eigenvalue problem \eqref{eq:local_eqn_inner} is then given by the linear combination of the basis elements with coefficients given by $c_k(g)$:
 \begin{equation}
 G(x) \simeq\sum_{k=0}^{M-1} c_k(g)\, \phi_k(x)\,.
\end{equation}
Note that our original problem would require an infinite set of basis elements, i.e. $M\rightarrow\infty$, and truncating allows us to determine a finite number of coefficients $\alpha_n$ of the series $A(g)$, to some precision (which can be checked by varying the value of the truncation $M$).

We can now expand the vector $\textbf{c}(g)$ for small $g$,
\begin{equation}
    \textbf{c}(g) = \sum_{n=0}^\infty g^n \,\textbf{c}_n\,=\; \sum_{n=0}^\infty g^n \left(c_{0,n},\cdots,c_{M-1,n}\right)^T\,,
\end{equation}
where we defined the expansion of the components of the vector \eqref{eq:appB-vector-def} as 
\begin{equation}
    c_\ell(g)=\sum_{k=0}^{\infty}\,g^n\,c_{\ell,n}\,,\;\ell=0,\cdots,M-1\,.
\end{equation}
Substituting the series of $\textbf{c}(g)$ into \eqref{eq:matrix_inner} and expanding for small $g$, we obtain the following set of matrix equations at each order of $g$
\begin{equation}
\label{eq:g_expanded_mat_eqns}
\begin{aligned}
      (\textbf{H} -\alpha_0 \textbf{W})\,\textbf{c}_{0}&=0,\\
    (\textbf{H}-\alpha_0\textbf{W})\,\textbf{c}_n &=-\textbf{V}\,\textbf{c}_{n-1} +\sum^{n}_{k=1}\alpha_k \textbf{W} \,\textbf{c}_{n-k}, \quad n\geq 1.
\end{aligned}
\end{equation}
By definition $\textbf{W}$ is a symmetric matrix; with the choice of basis made below it is also positive definite. We can therefore use the Cholesky decomposition to write it as 
\begin{equation}
    \textbf{W} = \textbf{L} \textbf{L}^T
\end{equation}
where $\textbf{L}$ is a lower triangular matrix with real and positive diagonal entries.
Next, we define the following 
\begin{equation}
    \widetilde{\textbf{H}} = \textbf{L}^{-1}\textbf{H} (\textbf{L}^{T})^{-1}, \quad \ \widetilde{\textbf{V}} = \textbf{L}^{-1}\textbf{V} (\textbf{L}^{T})^{-1}, 
\end{equation}
and 
\begin{equation}
    \widetilde{\textbf{c}} = \textbf{L}^{^T} \textbf{c}.
\end{equation}
In terms of the tilde quantities, the equations \eqref{eq:g_expanded_mat_eqns} can be written as 
\begin{equation} \label{eq:tilde_mat_eqn}
    \begin{aligned}
      (\widetilde{\textbf{H}} -\alpha_0 )\,\widetilde{\textbf{c}}_0&=0\,,\\
    (\widetilde{\textbf{H}}-\alpha_0)\,\widetilde{\textbf{c}}_n &=-\widetilde{\textbf{V}} \,\widetilde{\textbf{c}}_{n-1} +\sum^{n}_{k=1}\alpha_k  \,\widetilde{\textbf{c}}_{n-k}, \quad n\geq 1\,.
\end{aligned}
\end{equation}
To calculate $\alpha_n$ we first note that $\widetilde{\textbf{c}}_0$ must be an eigenvector of $\widetilde{\textbf{H}}$. For this reason, we now label the vector with a superscript $\widetilde{\textbf{c}}_0^{(i)}$ meaning that
\begin{equation}
    \widetilde{\textbf{H}} \, \widetilde{\textbf{c}}^{(i)}_0 =\alpha_0^{(i)} \,\widetilde{\textbf{c}}_0^{(i)}
\end{equation}
where the $i$ denotes that this is the $i^{th}$ eigenvector of $\widetilde{\textbf{H}}$ and $\alpha_0^{(i)}$ is the $i^{th}$ eigenvalue. 
We fix the normalisation of this eigenvector by demanding that 
\begin{equation}
    \left(\widetilde{\textbf{c}}_0^{(i)}\right)^T\widetilde{\textbf{c}}_0^{(i)} =1.
\end{equation}
Furthermore, we denote by $\widetilde{\textbf{c}}^{(i)}_n$ for $n\geq 1$, the solutions to the recursion \eqref{eq:tilde_mat_eqn} starting with the $i^{th}$ eigenvector $\widetilde{\textbf{c}}_0^{(i)}$. The equations \eqref{eq:tilde_mat_eqn} then imply that if $\widetilde{\textbf{c}}^{(i)}_m$ is a solution to the second equation for a given fixed value of $m$, then so is  $\widetilde{\textbf{c}}^{(i)}_m +\beta \widetilde{\textbf{c}}_0^{(i)} $, for any $\beta \in \mathbb{C}$. We fix this ambiguity by demanding that 
\begin{equation}
       \left(\widetilde{\textbf{c}}_0^{(i)}\right)^T\widetilde{\textbf{c}}_n^{(i)} =0, \qquad n\ge1.
\end{equation}

Given that $\widetilde{\textbf{H}}$ is positive definite and symmetric, we can order its eigenvalues, and taking $i=0$ to be the smallest eigenvalue that corresponds to the hydrodynamic mode, we therefore focus on this case. If we take the inner product of the second line of the recursion relation
\eqref{eq:tilde_mat_eqn} with $\widetilde{\textbf{c}}_0^{(i)}$ for $n \geq 1$ leads to the following equations
\begin{equation}
\begin{aligned}\label{eq:rec_rel_nums}
    \alpha_n^{(0)} &=\left(\widetilde{\textbf{c}}^{(0)}_{0}\right)^T\widetilde{\textbf{V}} \,\widetilde{\textbf{c}}^{(0)}_{n-1} \,, \\
    \left(\widetilde{\textbf{c}}_0^{(i)}\right)^T \,\widetilde{\textbf{c}}^{(0)}_n &=  \frac{1}{\alpha_0^{(i)}-\alpha_0^{(0)}}\left(-\left(\widetilde{\textbf{c}}^{(i)}_{0}\right)^T\widetilde{\textbf{V}} \,\widetilde{\textbf{c}}^{(0)}_{n-1}+\sum_{k=1}^n \alpha_k^{(0)}\left(\widetilde{\textbf{c}}^{(i)}_{0}\right)^T \widetilde{\textbf{c}}^{(0)}_{n-k}\right), \quad i\geq 1 \,.
\end{aligned}
\end{equation}
This gives a recursive way of calculating the quantities $\alpha_n^{(0)}$ and $\left(\widetilde{\textbf{c}}_0^{(i)}\right)^T \,\widetilde{\textbf{c}}^{(0)}_n $. The quantities in the recursion involving $\left(\widetilde{\textbf{c}}^{(0)}_{0}\right)^T\widetilde{\textbf{V}} \,\widetilde{\textbf{c}}^{(0)}_{n-1} $ can be computed by inserting the complete set of eigenvectors $\widetilde{\textbf{c}}_0^{(j)}$
\begin{equation}\label{eq:V_decomp}
    \left(\widetilde{\textbf{c}}^{(i)}_{0}\right)^T\widetilde{\textbf{V}} \,\widetilde{\textbf{c}}^{(0)}_{n-1} = \sum_{j=0}^{M-1}  \left[\left(\widetilde{\textbf{c}}^{(i)}_{0}\right)^T\widetilde{\textbf{V}} \,\widetilde{\textbf{c}}^{(j)}_{0}\right]    \left[\left(\widetilde{\textbf{c}}^{(j)}_{0}\right)^T\,\widetilde{\textbf{c}}^{(0)}_{n-1}\right]\,.
\end{equation}
Then given that we have computed the matrices $\textbf{H}$, $\textbf{V}$ and $\textbf{W}$ to some size $M$, the process of calculating $\alpha_n^{(0)}$ is as follows:
\begin{enumerate}
    \item Compute \textbf{L} using the Cholesky decomposition.
    \item Use \textbf{L} to compute  $\widetilde{\textbf{H}}$ and $\widetilde{\textbf{V}}$.
    \item Compute the eigenvalues $\alpha_0^{(i)}$ and unit eigenvectors $\widetilde{\textbf{c}}_0^{(i)}$ of $\widetilde{\textbf{H}}$. 
    \item Say that we have computed up to $\alpha_{n-1}^{(0)}$ and $\left(\widetilde{\textbf{c}}_0^{(i)}\right)^T \,\widetilde{\textbf{c}}^{(0)}_{n-1}$, compute the quantity  $\left(\widetilde{\textbf{c}}^{(i)}_{0}\right)^T\widetilde{\textbf{V}} \,\widetilde{\textbf{c}}^{(0)}_{n-1}$ using \eqref{eq:V_decomp}, the eigenvectors $\widetilde{\textbf{c}}_0^{(i)}$ and the inner product $\left(\widetilde{\textbf{c}}_0^{(i)}\right)^T \,\widetilde{\textbf{c}}^{(0)}_{n-1}$.
    \item Use  $\left(\widetilde{\textbf{c}}^{(i)}_{0}\right)^T\widetilde{\textbf{V}} \,\widetilde{\textbf{c}}^{(0)}_{n-1}$ and equation \eqref{eq:rec_rel_nums} to compute $\alpha_n^{(0)}$.
    \item  Use the newly computed $\alpha_n^{(0)}$ in \eqref{eq:rec_rel_nums} to compute  $\left(\widetilde{\textbf{c}}_0^{(i)}\right)^T \,\widetilde{\textbf{c}}^{(0)}_{n}$.
    \item Return to step $4$.
\end{enumerate}

\subsection{Basis functions}

In this subsection, we discuss the basis functions $\phi_i(x)$ we use in the process described in the previous subsection. We also use their properties to compute the matrices $\textbf{H}$, $\textbf{W}$ and $\textbf{V}$. The functions we pick for the spectral method discussed in subsection \ref{ap:general_meth} are
\begin{equation}
    \phi_n(x) = x\, \rme^{-\beta x/2} \,L_n^{(1)}(\beta x).
\end{equation}
where $L_n^{(1)}$ is the $n^{th}$ generalised Laguerre polynomial. These functions obey the boundary condition at $x =0$ and $x =\infty$ while keeping the matrices $\textbf{H}$, $\textbf{V}$ and $\textbf{W}$ relatively simple. The parameter $\beta$ can be chosen and different choices of $\beta$ should not change the results. 
To calculate $\textbf{H}$, $\textbf{V}$ and $\textbf{W}$ in \eqref{eq:appB-matrices-def} we use four key properties of the generalised Laguerre polynomials. 
\begin{equation}\label{eq:lag_props}
\begin{aligned}
     &\int_0^\infty L_n^{(\gamma)}(x)L_m^{(\gamma)}(x) \rme^{-x} x^\gamma = \frac{\Gamma(n + \gamma+1)}{n!} \delta_{n,m}\\
  & xL_{k}^{(\gamma)}(x) = -(k+\gamma)L_{k-1}^{(\gamma)}(x)+ (2k+1+\gamma)L_{k}^{(\gamma)}(x)-(k+1)L_{k+1}^{(\gamma)}(x)\\
  &\frac{d}{dx}\left(\rme^{-x} x^\gamma L_k^{(\gamma)}(x)\right) = (k+1)\rme^{-x}x^{\gamma-1}L_{k+1}^{(\gamma-1)}(x)\\
  & x L_i^{(\gamma+1)}(x) = -(i+1)L_{i+1}^{(\gamma)}(x) +(i +\gamma +1)L_{i}^{(\gamma)}(x)
\end{aligned}
\end{equation}
$\textbf{W}$ is easily computed using the first property: 
\begin{equation}
     W_{i, j}= \int_0^\infty x \rme^{-\beta x} L_i^{(1)}(\beta x)L_j^{(1)}( \beta x)\, dx \,=\, \frac{j+1}{\beta^2}\delta_{i, j}.
\end{equation}
The matrix elements of $\textbf{V}$ are given by
\begin{equation}
     V_{i, j} = \int_0^\infty x^2 \rme^{-\beta x} L_i^{(1)}(\beta x)L_j^{(1)}( \beta x)\, dx.
\end{equation}
Using the first and second property of \eqref{eq:lag_props} we have:
\begin{equation}
    \begin{aligned}
        &V_{i,i}= \frac{2}{\beta^3}(i+1)^2,\\
        &V_{i,i+1}=V_{i+1,i}= -\frac{(i+1)(i+2)}{\beta^3},\\
        &V_{i,j}=0, \quad|i-j|>1.
    \end{aligned}
\end{equation}
Thus $\textbf{V}$ is a tridiagonal matrix. Finally, we compute $\textbf{H}$. First, we use the third and fourth properties of \eqref{eq:lag_props} to deduce
\begin{equation}
    \phi_i'(x) = \frac{1}{2}(i +1)\rme^{-\beta x/2}\left(L_{i+1}^{(0)}(\beta x) +L_{i}^{(0)}(\beta x) \right).
\end{equation}
Then using the first property of \eqref{eq:lag_props} with $\gamma =0$, we have
\begin{equation}\label{eq:ham_der}
    \int_0^\infty\phi_i'(x) \phi'_j(x) \,dx= \frac{(i+1)(j+1)}{4\beta}\left( \delta_{i, j+1}+\delta_{i+1, j}+2\delta_{i, j}\right)\,.
\end{equation}
If we denote
\begin{equation}
  \beta^4X_{i,j}=  \int_0^\infty x\phi_i(x)\phi_j(x)   \,dx\,,
\end{equation}
then using the second property of \eqref{eq:lag_props}, we obtain
\begin{equation}
\begin{aligned}
      & X_{i,i}=6(i+1)^3,\\
       &X_{i,i+1}=X_{i+1, i}=-2(i+1)(i+2)(2i+3),\\
       &X_{i,i+2}=X_{i+2, i}= (i+1)(i+2)(i+3),\\
       &X_{i,j}=0, \quad|i-j|>2.
\end{aligned}
\end{equation}
Adding this result to equation \eqref{eq:ham_der}, we obtain 
\begin{equation}
\begin{aligned}
      &H_{i,i} = \frac{(i+1)^2}{2\beta} + \frac{6(i+1)^3}{\beta^4},\\
         & H_{i,i+1}=H_{i+1,i} = \frac{(i+1)(i+2)}{4\beta} - \frac{2(i+1)(i+2)(2i+3)}{\beta^4},\\
              &H_{i,i+2} =H_{i+2,i} = \frac{ (i+1)(i+2)(i+3)}{\beta^4},\\
               &H_{i,j}=0, \quad|i-j|>2.
\end{aligned}
\end{equation}
\bibliographystyle{ytphys}
\bibliography{hydro-biblio}

\end{document}